\documentclass[twocolumn]{aastex631}
\usepackage{amssymb,amsmath}

\shorttitle{High-$z$ galaxy constraints on cosmology and star formation}
\shortauthors{Sahl\'en \& Zackrisson}
\graphicspath{{./}{figures/}}

\begin{document}

\title{Galaxy population constraints on cosmology and star formation in the early Universe}

\correspondingauthor{Martin Sahl\'en}
\email{msahlen@msahlen.net}

\author[0000-0003-0973-4804]{Martin Sahl\'en}
\affiliation{Theoretical astrophysics, Department of Physics and Astronomy, Uppsala University, Box 515, SE-751 20 Uppsala, Sweden}
\affiliation{Swedish Collegium for Advanced Study, Thunbergsv\"{a}gen 2, SE-752 38 Uppsala, Sweden}

\author[0000-0003-1096-2636]{Erik Zackrisson}
\affiliation{Observational astrophysics, Department of Physics and Astronomy, Uppsala University, Box 515, SE-751 20 Uppsala, Sweden}



\begin{abstract}
We present the first post-cosmic-microwave-background early-Universe observational constraints on $\sigma_8$, $\Omega_{\rm m}$, mean galaxy star-forming efficiency and galaxy UV magnitude scatter at redshifts $z = 4-10$. 
We perform a simultaneous 11-parameter cosmology and star-formation physics fit using the new code GalaxyMC, with redshift $z>4$ galaxy UV luminosity and correlation function data. Consistent with previous studies, we find evidence for redshift-independent star formation physics, regulated by halo assembly. For a flat $\Lambda$CDM universe with a low-redshift Hubble constant and a Type Ia supernovae $\Omega_{\rm m}$ prior, we constrain $\sigma_8 = 0.81 \pm 0.03$, and a mean star-forming efficiency peaking at $\log_{10} {\rm SFE} = -[(0.09 \pm 0.20) + (0.58 \pm 0.29) \times \log_{10} (1+z)]$ for halo mass $\log_{10} M_{\rm p} / h^{-1} M_{\sun} = 11.48 \pm 0.09$. The suppression of star formation due to feedback is given by a double power law in halo mass with indices $\alpha = 0.56 \pm 0.08, \beta = -1.03 \pm 0.07$. The scatter in galaxy UV magnitude for fixed halo mass is $\sigma_M = 0.56 \pm 0.08$. Without a prior on $\Omega_{\rm m}$ we obtain $\sigma_8 = 0.78 \pm 0.06$, $\Omega_{\rm m} = 0.33 \pm 0.07$ and at most $1\sigma$ differences in all other parameter values.  Our best-fit galaxy luminosity functions yield a reionization optical depth $\tau \approx 0.048$, consistent with the {\it Planck} 2018 value.
\end{abstract}


\section{INTRODUCTION}
The final frontier of galaxy studies, the first generations of galaxies a few hundred million years after the Big Bang, has been opening up over the last decade \citep{Finkelstein16}. Large and growing samples of tens of thousands of galaxies at high redshifts $z = 4$--$12$ are now well-established and increasingly characterised in terms of luminosity and correlation functions \citep[e.g.,][]{2015ApJ...803...34B,Finkelstein2015,McLeod2016,Morishita2018,Oesch2018,Livermore2018,Ishigaki18,Hatfield2018,Harikane2018,Stefanon2019,Bowler2020,RojasRuiz2020}. This is thanks in particular to the successful application of the Lyman-break dropout selection technique and magnification by galaxy cluster gravitational lensing to optical-NIR surveys with telescopes like {\it Hubble}, VISTA and Subaru. With the advent of the {\it James Webb Space Telescope (JWST)} and {\it EUCLID}, new breakthroughs in surveying and understanding galaxies at $6 \lesssim z \lesssim 15$ can be expected.  It is therefore timely to consider how such samples may be used to constrain models of both cosmology and galaxy/star formation. 

The observed distribution of high-redshift galaxies depends on a number of components: the primordial density perturbations; the cosmic expansion and linear growth histories; the properties of dark matter, dark energy or other exotic constituents in the early Universe; the history of non-linear halo assembly; the impact of external feedback on galaxy formation; the statistical relation between halo mass and observed apparent galaxy magnitude including dust extinction; and the relevant survey selection. The abundance of high-redshift galaxies has already been used to place interesting constraints on non-standard dark matter models \citep{2013MNRAS.435L..53P,2014MNRAS.442.1597S,2016ApJ...825L...1M,2017PhRvD..95h3512C,2017ApJ...836...61M}, early dark energy \citep{2020ApJ...900..108M} and primordial non-gaussianity on small scales \citep{2021JCAP...01..010S}. High-redshift galaxies are thus an interesting novel probe sensitive to a variety of new early-Universe physics that could help reconcile the observational tension between late-Universe and early-Universe measurements of the Hubble constant $H_0$ \citep{2021arXiv210301183D} and the matter power spectrum normalization $\sigma_8$ \citep{2021A&A...646A.140H}.

The luminosity functions (LFs) and correlation functions (CFs) of high-redshift galaxies have been investigated at different redshifts to understand their connection to the underlying distribution of dark matter and its growth and evolution over time. It has been found that the galaxy population appears to evolve predominantly with the population of dark matter halos, implying that the relevant physics of star formation remains relatively constant over time \citep[e.g.,][]{Trenti2010,Behroozi2013,Dayal2013,Tacchella2013,Mason2015,Sun2016,Tacchella2018,Harikane2018,Yung2019a,Behroozi2019}. The population of high-redshift galaxies are likely the dominant sources for reionization, and their formation can also be suppressed due to feedback from the reionization process \citep{2018PhR...780....1D}.

This is the first in a series of papers that will investigate cosmology, structure and star formation, and reionization in the early Universe, using the new code GalaxyMC\footnote{\url{http://galaxymc.space}}. Later papers will investigate additional models, probes and data, including 21cm Cosmic Dawn / Epoch of Reionization data. Here, we present the basic modelling and demonstrate parameter constraints within the flat $\Lambda$CDM cosmological model using the observed galaxy UV LF at redshifts $z = 4 - 10$ and the galaxy CF at $z = 4 - 7$.  We employ the same general, semi-analytic approach as in several recent studies, whereby the star-forming efficiency is parametrized and the parameter values directly fitted to observations \citep[see, e.g.,][]{Moster2010,Tacchella2013,Behroozi2013,Sun2016,Furlanetto2017,Park2019,2020MNRAS.491.3891P,Mirocha2020b}. However, we do not directly model halo assembly histories, but subsume their effect into an effective galaxy UV magnitude scatter for fixed halo mass. Our model is motivated by compromise between sufficient physical detail in relation to observational precision and computational tractability. Compared to earlier works, we also introduce modeling to account for dependence on cosmological model parameters, and employ multiply parallelized numerical routines to enable Monte Carlo Markov Chain (MCMC) exploration of both star formation and cosmological parameter space. Thanks to the attained efficiency, we can include the relevant galaxy UV LF data in full across all redshifts. We also include a measurement of the galaxy correlation function as an additional constraint. This allows us to -- for the first time -- place observational large-scale structure constraints on the current normalization of the matter power spectrum, $\sigma_8$, and mean matter density $\Omega_{\rm m}$ based on $z>4$ post-cosmic-microwave-background data, jointly with parameters of star formation physics. We also compare the implications of our best-fit galaxy UV LF for the global history of reionization with current observational constraints. 

Logarithms are base 10 unless otherwise specified.

\section{DATA}
We use the galaxy luminosity function determinations derived from $\sim 10\,000$ Lyman-break selected high-redshift galaxy candidates across a redshift range $z \sim 4 - 10$ \citep[][\bf B15]{2015ApJ...803...34B}. This data set is a large, to good approximation magnitude-limited and homogeneously analyzed sample that has been extensively used in the literature. We use the derived galaxy LF, Table 5 in \cite{2015ApJ...803...34B}. We discuss completeness and other aspects further in Sect.~\ref{sec:syst}. 

We also include measurements of mean galaxy magnitudes as a function of halo mass, determined from the galaxy CF \citep[][\bf H18]{Harikane2018}. We use a subset of the data points in their Fig.~15, chosen to give a maximal lever arm in halo mass and redshift, but otherwise minimal to allow the LF data to drive the results. Specifically, we use the two data points with smallest and largest halo mass at $z \sim 4$, and the data point with smallest halo mass at $z \sim 7$, as a constraint on the mean UV magnitude of halos as a function of \emph{dark-matter-only} halo mass $M_{200}$ in the fiducial cosmology of \cite{Harikane2018}. We propagate the published halo mass uncertainty to a magnitude uncertainty, and take care to convert between the halo mass definition used in \cite{Harikane2018} and that used in our model. For the latter, we assume an NFW profile halo concentration parameter $c = 3.1$ \citep{2015MNRAS.452.1217C,2016MNRAS.462..893R}, consistent with assumptions in \cite{Harikane2018}. This also includes converting between a dark-matter-only mass and a dark-matter-plus-baryons mass. We compute the predicted magnitude in the fiducial cosmology of \cite{Harikane2018}. Hence, this is effectively an independent constraint on the star-forming efficiency. 
\vspace{15mm}

\section{MODEL}
\subsection{Galaxy luminosity function}
To model the observed distribution of galaxies, we employ a model combining a halo mass function with an effective mass--observable relation.

Our model to predict the comoving number density of galaxies at redshift $z$ with absolute magnitude $M_a \leq M \leq M_b$ is given by
\begin{eqnarray}
\label{eq:phi}
\Phi(M_a, M_b, z) & = & \epsilon_{\rm syst} \int_M \int_{M_{\rm min}(z)} W[M; M_a, M_b] 
\\
\nonumber
& & \times p[M | \langle M\rangle (M_{\rm h}, z)] n_h[M_{\rm h}, z] dM_{\rm h} dM, 
\end{eqnarray}
where $\epsilon_{\rm syst}$ describes fractional systematic uncertainty (e.g., cosmic variance, uncertainty in halo/sub-halo mass function, completeness, contamination), $M_{\rm min}(z)$ is the minimum halo mass allowing galaxy/star formation, $W[M; M_a, M_b]$ is a top-hat window function picking out absolute magnitudes $M_a \leq M \leq M_b$, $p[M | \langle M\rangle (M_{\rm h}, z)]$ is the probability of a galaxy having absolute magnitude $M$ given a mean galaxy magnitude $\langle M\rangle (M_{\rm h}, z)$,  $n_h(M_{\rm h},z)$ is the halo mass function. The integration variables are intrinsic, absolute magnitude $M$ and halo mass $M_{\rm h}$. 

\subsection{Background and density perturbations}
We assume a flat $\Lambda$CDM cosmology described by the present mean matter density $\Omega_{\rm m}$, baryonic matter density $\Omega_{\rm b}$, and Hubble constant $H_0$ (also denoted by $h = H_0 / 100$ km/s/Mpc), with adiabatic scalar primordial density perturbations that have a power-law power spectrum described by the present-time normalization $\sigma_8$ and scalar spectral index $n_{\rm s}$, and no higher-order statistical moments. We assume that cosmological neutrinos are massless, and that there are $N_{\rm eff} = 3.046$ effective relativistic species in the early Universe.

\subsection{Halo mass function} 
The differential number density of halos in a mass
interval ${\rm d}M$ about $M$ at redshift $z$ can be written as
\begin{equation}
n(M_{\rm h},z)\,{\rm d}M_{\rm h} = -F(\sigma)\,\frac{\rho_{\rm m}(z)}{M_{\rm h}\sigma(M_{\rm h},z)}\,
\frac{{\rm d}\sigma(M_{\rm h},z)}{{\rm d}M_{\rm h}}\,{\rm d}M_{\rm h}\,,
\end{equation}
where $\sigma(M_{\rm h},z)$ is the dispersion of the density field at some
comoving scale $R=(3M_{\rm h}/4\pi\rho_{\rm m})^{1/3}$ and redshift $z$, and
$\rho_{\rm m}(z) = \rho_{\rm m}(z=0)(1+z)^3$ the matter density.

The halo mass function $F(\sigma)$ encodes the halo collapse statistics. We use the fit to Bolshoi and MultiDark $N$-body simulations in \cite{2016MNRAS.462..893R} [Eq. (25), Eq. (32), Table 3],  
calibrated on simulations up to $z = 9$. We use this fit up to $z = 10$, and for $10<z\leq 15$ linearly interpolate $\log F(\sigma)$ in redshift using our $z = 10$ value and the results in \cite{Yung20}, Appendix A. We expect that this determination is accurate to within $\sim 20\%$ for the relevant halo masses. The cited works use the virial overdensity at each redshift to define halo masses, so we accordingly let $M_{\rm h}$ correspond to the halo virial mass $M_{\rm vir}$. 
For simplicity, we assume that halos host a central galaxy only, since at high redshift the fraction of satellites is expected to be at most a few percent \citep{2016MNRAS.462..893R, 2018MNRAS.480.3177B}. This is also within the theoretical uncertainty of the halo mass function quoted above \citep{2016MNRAS.462..893R}.

\subsection{Minimum mass of galaxy formation}
Both internal and external processes can inhibit the possibility to form galaxies in the early Universe. We model internal feedback through the star-forming efficiency model in Sect.~\ref{sec:sfe} below. Both the HI cooling limit of dark matter halos and photo-suppression by the ultraviolet radiation field built up during reionization can also prevent the formation of galaxies. The data we use are not faint enough to be sensitive to these mechanisms, but for definiteness, we model the minimum halo mass due to HI cooling and UV background feedback using Eq.~(11) in \cite{2013MNRAS.432.3340S} with central redshift and duration of reionization $z_{\rm re} = 7.7, \Delta z_{\rm re}=0.5$ following the Planck 2018 results \citep{Planck18}, and a sound-crossing redshift interval $\Delta_{\rm sc} = 1.0$.

\subsection{Galaxy star-forming efficiency and luminosity} \label{sec:sfe}
We assume that galaxy UV luminosities follow $L = \kappa^{-1} \dot{m}_\star$ \citep{2014ARA&A..52..415M}, with the star-formation rate (SFR) given by 
\begin{equation}
\label{eq:mstardot}
	\dot{m}_\star = f_\star \frac{\Omega_{\rm b}}{\Omega_{\rm m}} \dot{M}_h \,,
\end{equation}
where $f_\star$ is the star-forming efficiency (SFE), the fraction of infalling baryonic matter that is converted to stars. The absolute UV magnitude $M = M_0 - 2.5\log_{10} L$, so that the mean magnitude for a galaxy is given by
\begin{widetext}
\begin{eqnarray}
\label{eq:massmag}
\langle M_{\rm UV} \rangle (M_{\rm h}, z) & = & M_0 - 2.5\left[ \log_{10} f_\star(M_{\rm h}, z) + \log_{10}\frac{\Omega_{\rm b}}{\Omega_{\rm m}} + \log_{10}\frac{\dot{M_{\rm h}}(M_{\rm h},z)}{M_\sun {\rm yr}^{-1}} - \log_{10} \frac{\kappa}{M_\sun\,{\rm yr}^{-1} / ({\rm ergs}\,{\rm s}^{-1}\,{\rm Hz}^{-1})} \right] 
\end{eqnarray}
\end{widetext}
where $f_\star(M_{\rm h}, z)$ is now the mean star-forming efficiency, and $\dot{M_{\rm h}}(M_{\rm h},z)$ is the mean mass accretion rate of halos. We use $\kappa = 1.15 \times 10^{-28} M_\sun\,{\rm yr}^{-1} / ({\rm ergs}\,{\rm s}^{-1}\,{\rm Hz}^{-1})$, calibrated for 1500 \r{A} dust-corrected rest-frame UV luminosity, assuming continuous mode star formation and a Salpeter stellar initial mass function \citep{2014ARA&A..52..415M}. We use the AB magnitude system \citep{1974ApJS...27...21O} so $M_0 = 51.6$. Note that this implies that additional effects from e.g. mergers, scatter in the initial mass function, scatter in filtering mass due to UV background feedback, etc. are subsumed into the statistical magnitude scatter.

The mass-magnitude relation in Eq.~(\ref{eq:massmag}) requires two input functions: the SFE $f_\star(M_{\rm h}, z)$, and the mass accretion rate (MAR) $\dot{M_{\rm h}}(M_{\rm h},z)$. We model the mean SFE as a double power-law distribution in $M_{\rm h}$:
\begin{equation}
    f_\star(M_{\rm h}, z) = \frac{N(1+z)^{\gamma_N}}{\left(\frac{M_{\rm h}}{M_{\rm p}}\right)^{\alpha(1+z)^{\gamma_\alpha}} + \left(\frac{M_{\rm h}}{M_{\rm p}}\right)^{-\beta(1+z)^{\gamma_\beta}}}\,.
\end{equation}
We assume throughout that $M_{\rm p}$ is given in units of $M_\sun/h$. This functional form is well-motivated based on earlier studies and allows the use of a single, general but compact function across all relevant redshifts and halo masses that implements internal feedback SFE suppression both at the low-mass (expected from supernova feedback) and high-mass (expected from feedback from active galactic nuclei) ends in a manner consistent with empirical studies \citep[see e.g.,][]{SunFurlanetto16,Furlanetto2017,Mirocha2020b}.

We model the MAR as
\begin{equation}
    \dot{M}_{\rm vir} = \frac{\sigma_8^{\rm sim}}{\sigma_8}\frac{H(z)}{H_{\rm sim}(z)} \dot{M}^{\rm sim}_{\rm vir} \,,
\end{equation}
where $H(z)$ and $H_{\rm sim}(z)$ are Hubble parameters, 
based on \cite{2015MNRAS.452.1217C} taking into account that the redshifts we consider are well within the matter-dominated regime [$\Omega_{\rm m}(z>4) \gtrsim 0.99$], such that the differences in growth rate for different choices of cosmological parameters are negligible. This model does not fully take into account the effect of changes in halo formation times when cosmological parameters are varied, but gives a good approximation to the mean MAR change. It should also be noted that for the ranges of cosmological parameter values considered in this work the variation in UV magnitudes is $\lesssim 0.1$ mag, and the overall impact on the galaxy LF dominated by the increase or decrease of halo densities.  The fiducial MAR $\dot{M}^{\rm sim}_{\rm vir}$ is given by a fit to the same $N$-body simulation used for the halo mass function \citep{2016MNRAS.462..893R},
\begin{eqnarray}
\label{eq:marsim}
\frac{\dot{M}^{\rm sim}_{\rm vir}}{h^{-1}\,M_\sun\,{\rm yr}^{-1}} & = & \eta(z) M_{\rm vir,12}^{\xi(z)}\frac{H_{\rm sim}(z)}{H^{\rm sim}_0}\,.
\end{eqnarray}
Here, 
\begin{eqnarray}
\log \eta(z) & = & 2.677 - 1.708a + 0.661a^2\,, \\
\xi(z) & = & 0.975 + 0.300a - 0.224a^2 \,.
\end{eqnarray}
The simulation has been performed with parameter values $H^{\rm sim}_0 = 67.8$ km/s/Mpc, $\Omega^{\rm sim}_{\rm m} = 0.307, \Omega^{\rm sim}_{\Lambda} = 0.693, \sigma_8^{\rm sim} = 0.823$. While the simulation data reaches up to $z = 9$, we have confirmed numerically that Eq.~(\ref{eq:marsim}) matches very well the MAR for $z = 9 - 15$ determined by assuming that the evolution of the halo mass function of \cite{Yung20} is due to mass accretion in individual halos. 
The model has a simple and natural physical interpretation: the accretion rates are regulated by the matter available for accretion from the background. If the degree of clustering is increased ($\sigma_8$) or the amount of matter is decreased ($\Omega_{\rm m}$), less matter will be available for accretion compared to the simulation baseline.

\subsection{Magnitude scatter}
The statistical scatter in galaxy UV magnitude is assumed to follow the distribution
\begin{eqnarray}
\nonumber 
    p(M_{\rm UV} | \langle M_{\rm UV} \rangle ) = \\  \frac{1}{\sqrt{2\pi}\sigma_M} \exp \left[ - \frac{(M_{\rm UV} - \langle M_{\rm UV} \rangle )^2}{2\sigma_{M}^2} \right] \,,
\end{eqnarray}
We allow the value of $\sigma_M$ to be determined by the data. 

\subsection{Dust extinction} 
We model dust extinction following \cite{2017PhRvD..95h3512C}, whereby the mean extinction at each UV magnitude $M_{\rm UV}$ and redshift $z$ is given by 
\begin{equation}
\langle A_{\rm UV} \rangle = 4.43 + 0.79\ln(10)\sigma^2_{\beta_{\rm UV}} + 1.99 \langle \beta_{\rm UV} \rangle
\end{equation}
with $\sigma_{\beta_{\rm UV}} = 0.34$ and $\langle \beta_{\rm UV}(M_{\rm UV}, z) \rangle$ approximated as in \cite{2017PhRvD..95h3512C} Eq.~(5-6).

\subsection{Fiducial model adjustment}
To compare our model predictions for the galaxy UV LF with the observational data, and allow cosmological parameters to be varied consistently, we rescale the model predictions to the fiducial cosmological model that was assumed in the derivation of the observed galaxy LF \citep{2015ApJ...803...34B}. When computing the predicted galaxy LFs from Eq.~(\ref{eq:phi}), we adjust galaxy absolute magnitudes as they would have appeared in the fiducial cosmology:
\begin{equation}
    M_{\rm UV}^{\rm fid} = M_{\rm UV} + 5 \log \left( \frac{d_{\rm L}(z)}{d^{\rm fid}_{\rm L}(z)} \right) \,,
\end{equation}
where $d_{\rm L}(z)$ is the luminosity distance. We also rescale the predicted number densities such that
\begin{equation}
    \Phi_{\rm fid} = \Phi \frac{dV_{\rm fid}/dz}{dV/dz}\,,
\end{equation}
where $dV/dz$ is the comoving volume element. Above, super- and subscripts ``fid'' refer to the fiducial cosmology. 

\subsection{Systematics} \label{sec:syst}
The estimated contamination in the B15 data set is low, with the contamination fraction approximately described by the relation $c(z) = [0.14(1+z)^2 - 1.7]/100$, based on \cite{2015ApJ...803...34B}. The galaxy LF data is already corrected for contamination, and hence we do not make any additional correction for this. The galaxy sample is estimated to be complete to at least $80\%$, and the observed number densities to have a $10\%$ uncertainty due to cosmic variance \citep{2015ApJ...803...34B}. The theoretical uncertainty in the halo/sub-halo mass function is of order $\sim 20\%$ \citep{2016MNRAS.462..893R}. Combining these estimates, we estimate an overall systematic uncertainty in galaxy number densities of around $15$-$20\%$. These uncertainties are subdominant to the statistical uncertainties on the galaxy LF, but we nonetheless include a normal-distribution prior $\epsilon_{\rm syst} = 1.00 \pm 0.15$.

Our assumed conversion between galaxy SFR and luminosity depends on assumptions about star formation dynamics, stellar initial mass function and metallicity, which could introduce a systematic bias. The uncertainty is expected to be within a factor of a few \citep{2014ARA&A..52..415M}, which is at worst comparable to the statistical uncertainty we find in the normalization $N$ of the SFE. 

\subsection{Likelihood and priors} 
We employ the measurements of the galaxy LF from \cite{2015ApJ...803...34B}, with corresponding normally-distributed uncertainties. We take care to appropriately truncate the probability density functions at zero where relevant.

\begin{figure*}[t]
\centering
\includegraphics[width=0.85\textwidth]{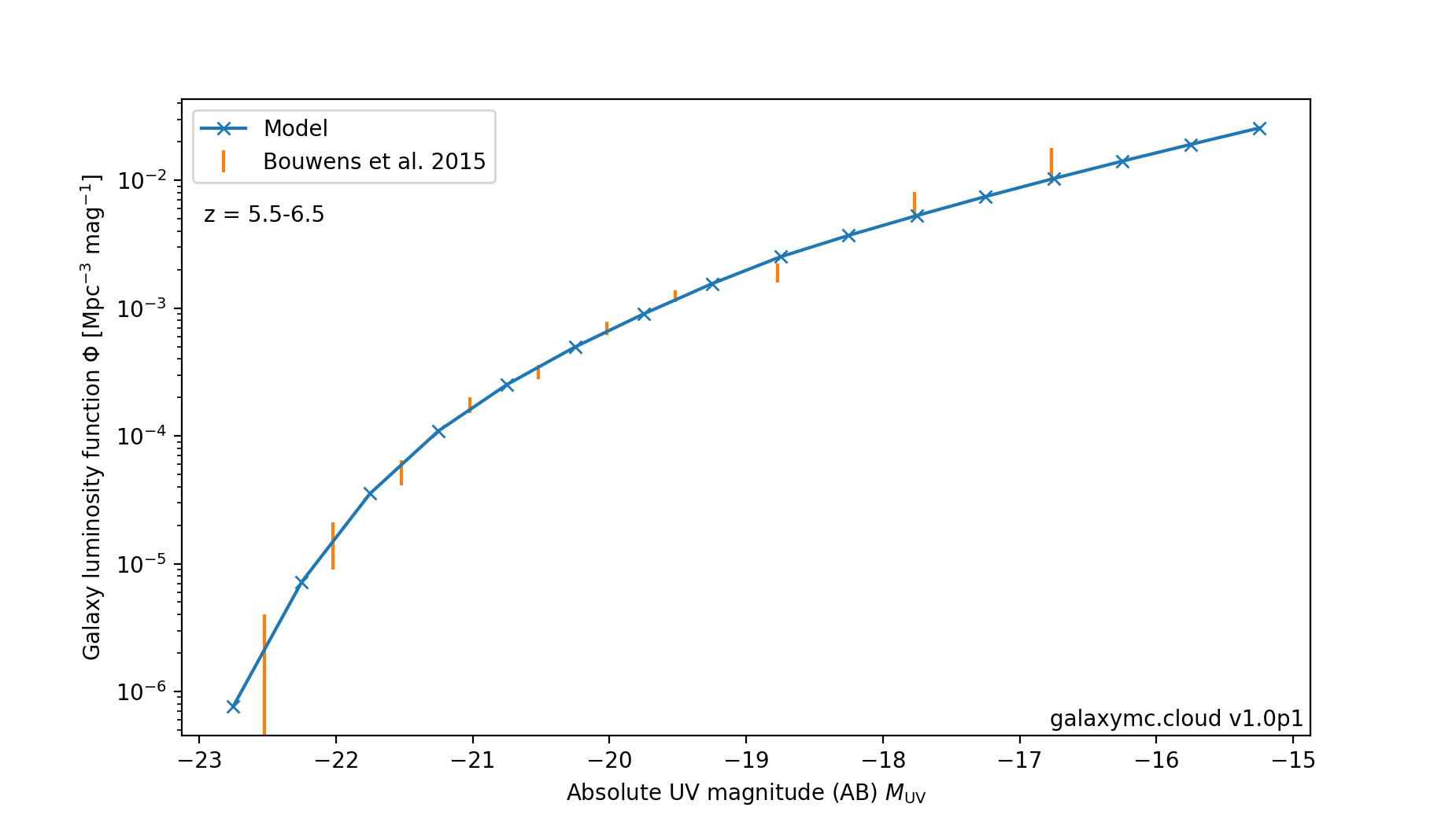}
\caption{The best-fit galaxy UV luminosity function $\Phi$ at $z = 6$ for prior combination I (indistinguishable from combination II), together with \citet[][B15]{2015ApJ...803...34B} data (with 68\% error bars), obtained from the online cloud computing application \url{galaxymc.cloud}. }
\label{fig:lumfunc}
\end{figure*}

We use non-restrictive, uniform priors on model parameters, unless otherwise stated. We enforce $N>0$ by employing instead the parameter $\log N$. We require $\alpha \geq 0$, $\beta \leq 0$ to avoid the artificial parameter degeneracy from index switching. The UV magnitude scatter $\sigma_M \geq 0.2$, in line with the theoretically expected minimum value possible due to variation in halo assembly based on analytical estimates and numerical simulations \citep{2018ApJ...856...81R, 2019ApJ...878..114R, 2020MNRAS.495.3602W}. We also restrict $h \leq 1$ to avoid an age of the Universe less than e.g. the ages of globular clusters \citep{2003Sci...299...65K}.

We employ the following external priors: 
\begin{widetext}
\begin{align}
& \mathbf{n_{\rm s}}\,{\rm\bf Planck} & & n_{\rm s} = 0.9649 & \text{\rm \cite{Planck18}} \,, \\
& \text{\rm\bf BBN} & & \Omega_{\rm b} h^2 = 0.0222 \pm 0.0005 & \text{\rm \cite{Planck18}} \,, \\ 
& {\rm\bf SNIa } & & \Omega_{\rm m} = 0.298 \pm 0.022 & \text{\rm \cite{2018ApJ...859..101S}} \,,\\
& \mathbf{H_0}\,{\rm\bf low-}\mathbf{z} && h = 0.7348 \pm 0.0166 & \text{\rm \cite{2018ApJ...855..136R}}\,, \\ 
& \mathbf{H_0}\,{\rm\bf Planck} && h = 0.674 \pm 0.005 & \text{\rm \cite{Planck18}} \,. \\
& {\rm\bf Syst.} & & \epsilon_{\rm syst} = 1.00 \pm 0.15 & \text{\rm See\,Sect.~\ref{sec:syst}.} 
\end{align}
\end{widetext}
For the Hubble constant, we test two different choices of prior, as indicated above. However, we do not combine the {\it Planck} Hubble constant prior with the corresponding {\it Planck} constraint on $\Omega_{\rm m}$, to distinguish the effect of changing $h$ alone. 

\section{COMPUTATION}
We perform a Monte Carlo Markov Chain exploration of the posterior probability distribution for the full set of 11 model parameters: $\{\Omega_{\rm b}, \Omega_{\rm m}, h, \sigma_8, \log N, \gamma_N, \log M_{\rm p}, \alpha, \beta, \sigma_M, \epsilon_{\rm syst}\}$. We test four different combinations of data and priors:
\begin{description}
\item [I] B15+H18+BBN+SNIa+$H_0$ (low-$z$)
\item [II] B15+H18+BBN+SNIa+$H_0$ ({\it Planck})
\item [III] B15+H18+BBN+$H_0$ (low-$z$)
\item [IV] B15+H18+BBN+SNIa.
\end{description}
The likelihood evaluations are performed using the hybrid FORTRAN-Python code GalaxyMC, a further development of the framework described in \cite{SahlenEqs}. An online cloud computing version of GalaxyMC, \url{galaxymc.cloud}, is publicly available\footnote{\url{http://galaxymc.cloud}}, and can be used to perform computations and visualizations of the galaxy LF model in this work. GalaxyMC performs cosmological calculations using CAMB \citep{2000ApJ...538..473L}, and numerical integrations using the state-of-the-art library Cuba \citep{2005CoPhC.168...78H, HAHN2016341}. About one hundred multi-dimensional integrations are performed at each likelihood evaluation. Thanks to parallelization, a likelihood evaluation takes on the order of a few seconds on a 10-core Intel Xeon E5 2630 v4 CPU at 2.20 GHz/core. The Monte Carlo exploration of the posterior is performed using the Emcee affine invariant ensemble sampling package \citep{2019JOSS....4.1864F} in Python, for which calculations are also parallelized.

Convergence of the Monte Carlo exploration of the posterior distribution is monitored by computing the auto-correlation length of the samples for each parameter using the Goodman-Weare estimator \citep{2010CAMCS...5...65G}, and requiring an effective sample size of at least 1100 samples to ensure robust credence regions \citep{2017ARA&A..55..213S}. We remove burn-in sections, confirm that a sampling acceptance rate $\sim 0.2 - 0.5$ is reached \citep{2013PASP..125..306F}, and visually inspect for good mixing and convergence. 

\section{RESULTS}

\begin{table*}[t]
     \centering
     \begin{tabular}{|c|c|c|c|c|}
     \hline
     Parameter & I. & II.  & III. & IV.  \\
     \hline
     $\Omega_{\rm m}$ & $\bf 0.30 \pm 0.02$ & $\bf 0.30 \pm 0.02$ & $0.33 \pm 0.07$ & $\bf 0.30 \pm 0.02$ \\
     \hline
     $h$ & $\bf 0.73 \pm 0.02$ & $\bf 0.674 \pm 0.005$ & $\bf 0.73 \pm 0.02$ & $0.74^{+0.15}_{-0.13}$  \\
     \hline
     $\sigma_8$ & $0.81 \pm 0.03$ & $0.84 \pm 0.03$ & $0.78 \pm 0.06$ & $0.82 \pm 0.06$ \\
     \hline
     $\log N$  & $0.21 \pm 0.20$ & $0.11 \pm 0.20$ & $0.13 \pm 0.26$ & $0.21 \pm 0.28$ \\
     \hline
     $\gamma_N$  & $-0.58 \pm 0.29$ & $-0.50 \pm 0.28$ & $-0.48 \pm 0.36$ & $-0.57 \pm 0.33$ \\ 
     \hline
     $\alpha$  & $0.55 \pm 0.07$ & $0.50 \pm 0.06$ & $0.55 \pm 0.07$ & $0.55^{+0.14}_{-0.12}$  \\ 
     \hline
     $\beta$ & $-1.03 \pm 0.07$ & $-1.02 \pm 0.07$ & $-1.06 \pm 0.09$ & $-1.03 \pm 0.07$ \\
     \hline
     $\log M_{\rm p}$ &  $11.48 \pm 0.09$ & $11.45 \pm 0.08$ & $11.49 \pm 0.09$ & $11.48 \pm 0.11$  \\
     \hline
     $\sigma_{M}$ & $0.56 \pm 0.08$ & $0.48 \pm 0.09$ & $0.57 \pm 0.08$ & $0.56^{+0.14}_{-0.20}$  \\
    \hline
     \end{tabular}
     \caption{Marginalized parameter constraints based on the \citet[][B15]{2015ApJ...803...34B} galaxy luminosity function, $z = 4 - 10$, and measurements of the galaxy correlation function from \citet[][H18]{Harikane2018}, $z = 4 - 7$. Four combinations of external priors on $h$ and $\Omega_{\rm m}$ are tested (I-IV). Constraints fully determined by external priors are marked in bold face.}
     \label{tab:lnresults}
\end{table*}

\subsection{Galaxy UV luminosity function} 

An example of the best-fit galaxy UV LF at $z = 6$ for case I, together with B15 data, is shown in Fig.~\ref{fig:lumfunc}. Our best-fit model predicts somewhat fewer galaxies at the faint end of the LF at $z\gtrsim 7$ compared to the reference LFs of \citet{Finkelstein16}. 
All fits (I-IV) reproduce the B15 data equally well, with reduced chi-square value $\chi^2_{\nu} \approx 1$. While acceptable within observational uncertainties, these findings likely reflect that our $z\lesssim 7$ data carries stronger statistical weight in the joint fit and also reaches fainter magnitudes, than our $z\gtrsim 7$ data, but could also be due to some additional redshift evolution in the mean SFE or UV magnitude scatter relative to $4\lesssim z \lesssim 7$ that our model does not fully capture. 

\subsection{Parameter inference}
\begin{figure}[t]
\includegraphics[width=\linewidth]{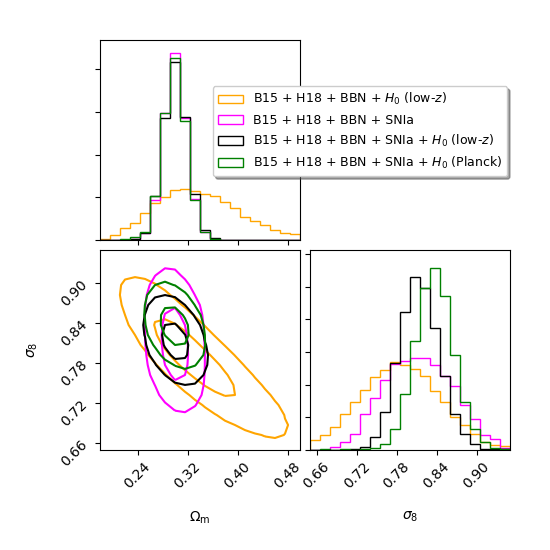}
\caption{Parameter constraints ($68\%$ and $95\%$ credence regions) on $\sigma_8$ and $\Omega_{\rm m}$ based on the \citet[][B15]{2015ApJ...803...34B} galaxy luminosity function, $z = 4 - 10$, and measurements of the galaxy correlation function from \citet[][H18]{Harikane2018}, $z = 4 - 7$. Four combinations of external priors on $h$ and $\Omega_{\rm m}$ are investigated (I-IV).}
\label{fig:corners8om}
\end{figure}

\begin{figure*}[t]
\includegraphics[width=\textwidth]{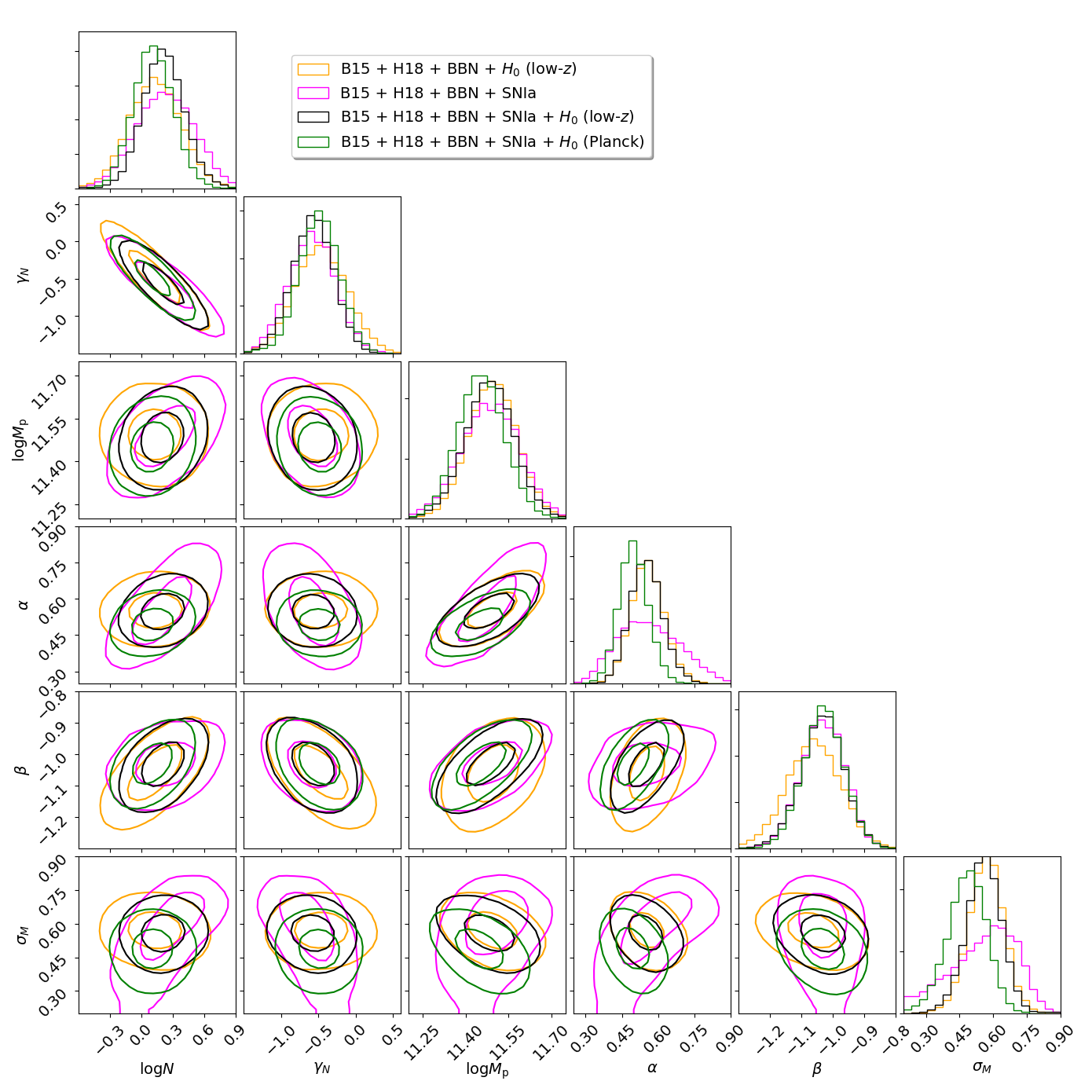}
\caption{Parameter constraints ($68\%$ and $95\%$ credence regions) on mean SFE model parameters based on the \citet[][B15]{2015ApJ...803...34B} galaxy luminosity function, $z = 4 - 10$, and measurements of the galaxy correlation function from \citet[][H18]{Harikane2018}, $z = 4 - 7$. Four combinations of external priors on $h$ and $\Omega_{\rm m}$ are investigated (I-IV).}
\label{fig:cornersfe}
\end{figure*}

\begin{figure*}[t]
\includegraphics[width=\textwidth]{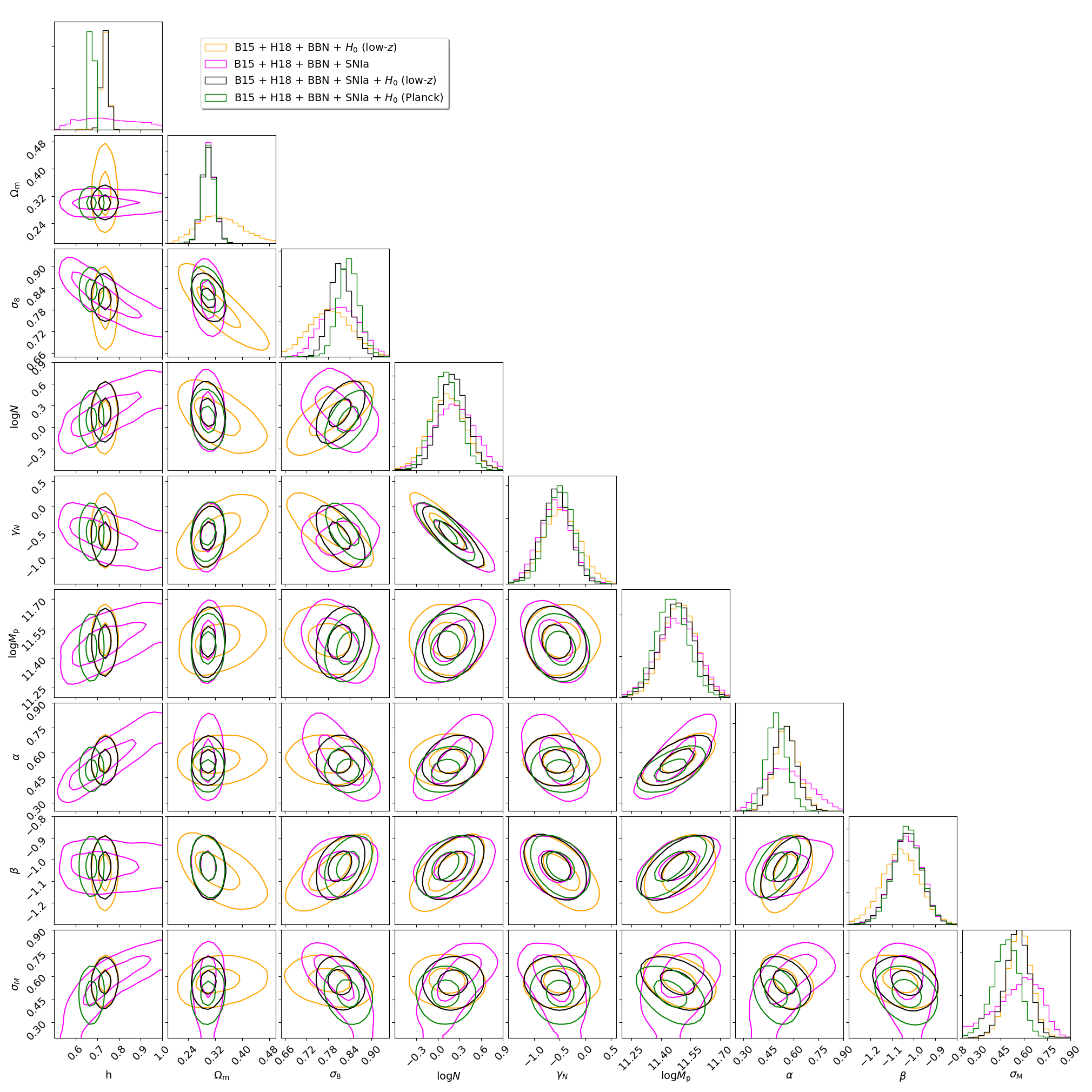}
\caption{Joint parameter constraints ($68\%$ and $95\%$ credence regions) on both cosmology and star formation based on the \citet[][B15]{2015ApJ...803...34B} galaxy luminosity function, $z = 4 - 10$, and measurements of the galaxy correlation function from \citet[][H18]{Harikane2018}, $z = 4 - 7$. Four combinations of external priors on $h$ and $\Omega_{\rm m}$ are investigated (I-IV).}
\label{fig:cornerall}
\end{figure*}

Parameter constraint contours are shown in Figs.~\ref{fig:corners8om}--\ref{fig:cornerall}. The marginalized parameter constraints are reported in Table~\ref{tab:lnresults}. All fits (I-IV) have reduced chi-square value $\chi^2_{\nu} \approx 1$.

We find that the inclusion of a redshift dependence in the SFE mass scaling (via parameters $\gamma_\alpha$ and $\gamma_\beta$) is not preferred by the data. Hence, in the following we restrict our analysis to the case where $\gamma_\alpha = \gamma_\beta = 0$. This is consistent with the findings in e.g.~\cite{Harikane2018, Tacchella2018}.

The inferred parameter values are clearly consistent in all the investigated cases, albeit with marginal tension in some cases. We find no evidence for a deviation from low-redshift cosmology across $z = 4$--$10$. The constraints on $\sigma_8$ are all rather tight and not very sensitive to conservative assumptions about $\Omega_{\rm m}$ and $h$ (in case III, IV), reflecting the sensitivity of halo formation to this parameter. The credence intervals for $\sigma_8$ are all consistent with current measurements from the cosmic microwave background \citep[CMB,][]{Planck18}, cluster abundance \citep[e.g.,][]{2019MNRAS.489..401Z} and weak lensing \citep{2021A&A...646A.140H}, but at least marginally in tension with the low-redshift measurements, especially for the case of a {\it Planck} Hubble constant prior (II). As suggested by the inverse degeneracy between $\sigma_8$ and $\Omega_{\rm m}$ seen in Fig.~\ref{fig:corners8om}, this tension could be relieved by a slightly larger value of $\Omega_{\rm m}$.

When $\Omega_{\rm m}$ is allowed to vary freely (III), it is still possible to obtain relatively good constraints on both $\sigma_8$ and $\Omega_{\rm m}$ (see Table~\ref{tab:lnresults}) thanks to the fortuitous orientation of their degeneracy in Fig.~\ref{fig:corners8om}.

When the Hubble constant $h$ is allowed to vary freely (IV, with $h\leq 1$), a weak constraint can be placed on its value (see Table~\ref{tab:lnresults}), with preferred values close to those of the low-$z$ Hubble constant prior (I). As a consequence, the preferred values of $\sigma_8$ are very similar in these two cases. 

The SFE normalization $N \sim 1.3-1.6$, consistent with the results in \cite{Harikane2018}. The redshift evolution of the SFE normalization, $\gamma_N$, is quite robustly inferred as negative, but is degenerate with $\log N$: a larger value of $\log N$ is compensated by a stronger negative redshift evolution. 
Our inferred values of the high-mass and low-mass SFE slopes $\alpha$, $\beta$ are consistent with the values found in other recent works \citep[e.g.,][cf. Table~\ref{tab:lnresults}]{Harikane2018,2020MNRAS.498.2645M}, although there defined relative to different peak halo masses $M_{\rm p}$, and despite our  inclusion of a UV magnitude scatter. The consistency is strengthened when considering their choices of $M_{\rm p}$ and the degeneracy between the two slope parameters $\alpha$, $\beta$ and $\log M_{\rm p}$ seen in Fig.~\ref{fig:cornersfe}.

\begin{figure*}[t]
\centering
\includegraphics[width=0.8\textwidth]{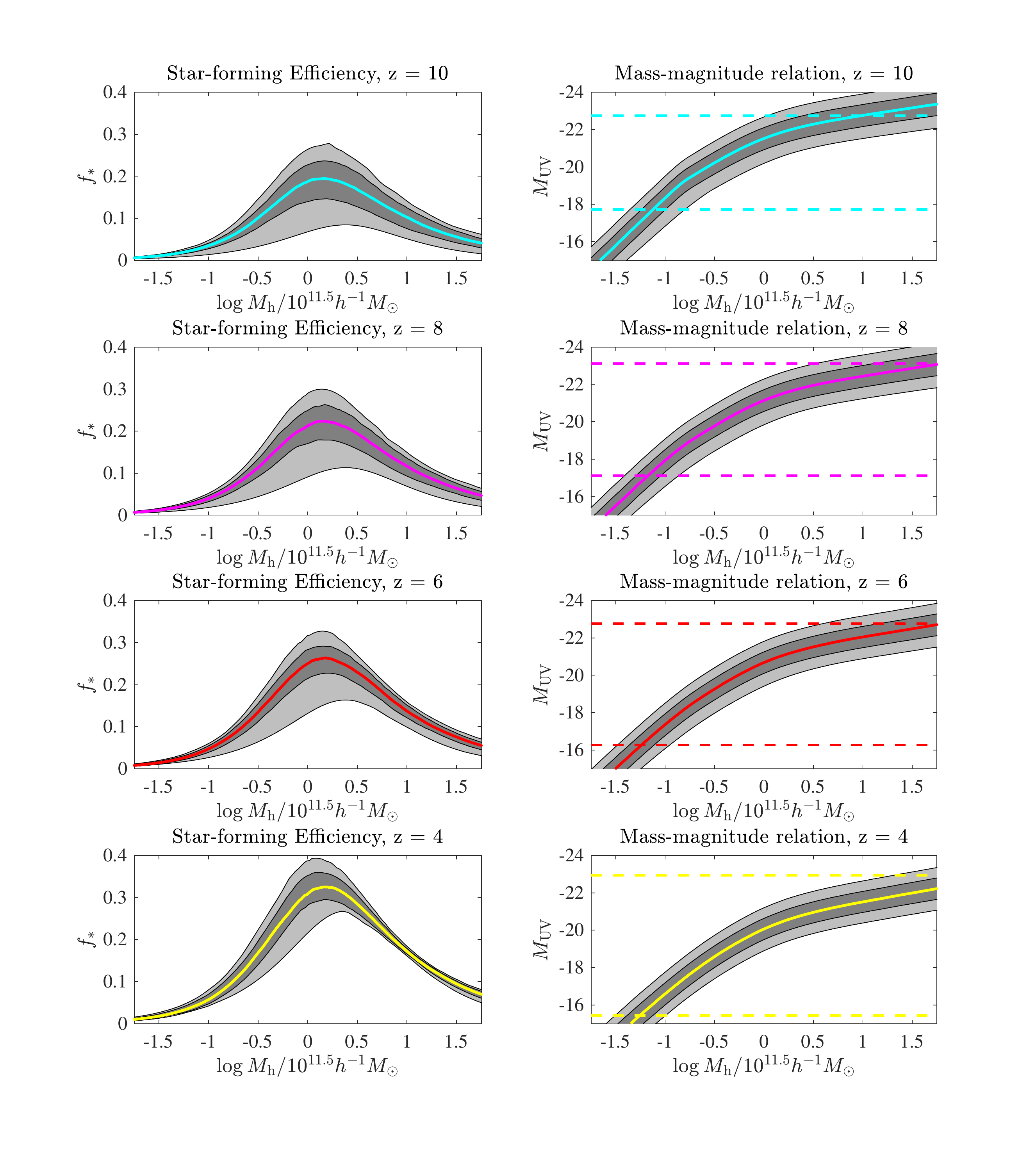}
\caption{Monte Carlo-derived star-formation quantities with 68\% and 95\% credence regions (dark and light gray shading) for prior combination I (indistinguishable from combination II). {\bf Left:} Star-forming efficiency as a function of halo mass. {\bf Right:} Halo mass - magnitude relation (\emph{with} dust extinction), including both statistical parameter uncertainties and the UV magnitude scatter.  The coloured dashed lines indicate the range of the \citet[][B15]{2015ApJ...803...34B} observational data at each redshift.}
\label{fig:sfemagrel}
\end{figure*}

While some degeneracies are present between SFE parameters (see Fig.~\ref{fig:cornersfe}), we see no significant changes in their marginalized parameter constraints between cases I-IV, except for the high-mass slope $\alpha$ and UV magnitude scatter $\sigma_{M}$. These show strong degeneracies with each other and with $h$ and $\Omega_{\rm m}$, as seen in Fig.~\ref{fig:cornerall}. This results in at least marginally significant shifts in $\alpha$ and $\sigma_{M}$ between low-$z$ and {\it Planck} Hubble constant priors (I, II), and only weak constraints when no prior is applied to $h$ or $\Omega_{\rm m}$ (III, IV). We can understand these degeneracies in the following way: changing the values of $h$, $\sigma_8$ and $\Omega_{\rm m}$ effectively rescales the whole galaxy LF \cite[cf. e.g., ][Fig.~8]{2020arXiv201000619V}, while scatter in UV magnitude for fixed halo mass ($\sigma_M$) predominantly affects the bright end of the LF \citep{2019ApJ...878..114R} where the effect of Malmquist bias is most prominent. The high-mass slope $\alpha$ also regulates the SFEs, hence galaxy magnitudes, at the bright end. In contrast, the SFE slope at the low-mass end ($\beta$) is quite robustly determined (see Fig.~\ref{fig:cornersfe} and Table~\ref{tab:lnresults}). At the fainter end of the galaxy LF, we move toward the rather flat bulk of the halo mass function. Galaxy abundance in this regime is not as sensitive to cosmological parameters as for high-mass halos. Since the halo mass function here is relatively flat in mass, scattering up and down in UV magnitude tend to cancel one another in the LF. Hence, the value of $\beta$ needed to reproduce observations should be relatively insensitive to $\sigma_M$ and cosmological parameters.

We determine a UV magnitude scatter $\sigma_M \sim 0.5 - 0.6$, in good agreement with the degree of scatter measured in other recent works \cite[e.g.,][]{2018ApJ...856...81R,Tacchella2018}. The degeneracies of $\sigma_M$ with other parameters are discussed above.

We marginalize all parameter constraints over $\Omega_{\rm b}$ and $\epsilon_{\rm syst}$, but have confirmed that their marginalized posteriors are consistent with their priors. 

\vspace{20mm}

\subsection{Implications for star formation and mass-magnitude relations}
As seen in Fig.~\ref{fig:sfemagrel}, the peak mean SFE is $ (19 \pm 5)\%$ at $z \sim 10$, rising to $(32 \pm 4)\%$ at $z \sim 4$. The mean SFEs and UV magnitudes as a function of halo mass show essentially no difference between a low-$z$ and a {\it Planck} Hubble constant prior (I, II), although the best-fit parameter values of the SFE model shift slightly. While we find clear evidence for the suppression of star formation due to feedback, one should bear in mind that the tight constraints on the mean SFE at small and large halo masses outside the range probed by observations (see dashed lines in right-hand column of Fig.~\ref{fig:sfemagrel}) are driven by the double power-law functional form assumed for the mean SFE.  As noted above, a UV magnitude scatter $\sigma_M \sim 0.5-0.6$ is preferred by the data. A $\sim 1\sigma$ shift is seen in the UV magnitude scatter towards smaller values in the {\it Planck} Hubble constant case (II).

\begin{figure*}[t]
\centering
\includegraphics[width=.45\textwidth]{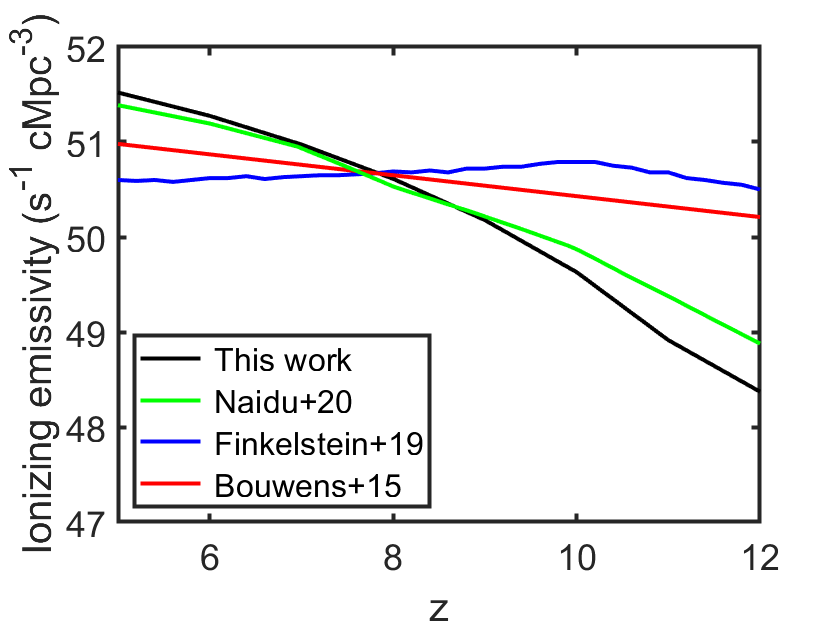}
\includegraphics[width=.45\textwidth]{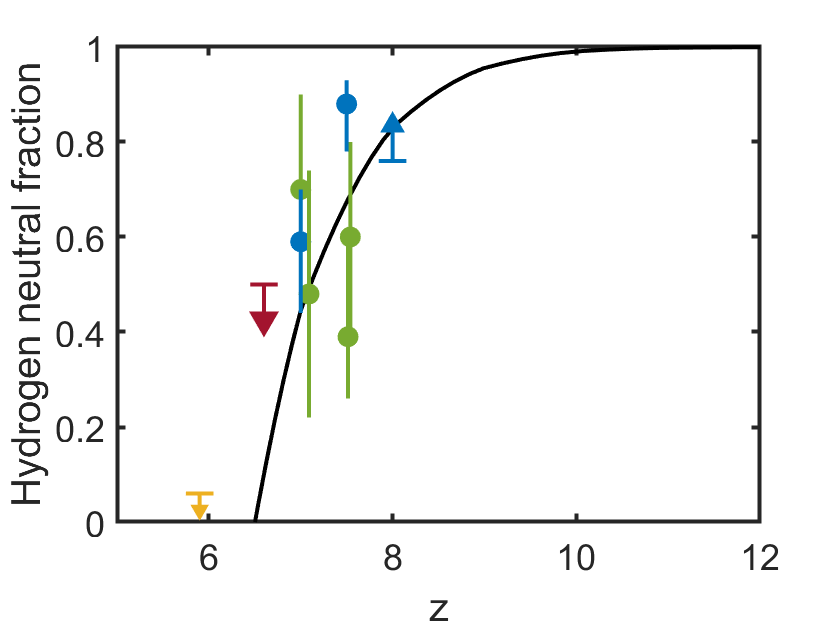}

\caption{{\bf Left:} The ionizing emissivity ($\dot{N}_\mathrm{ion}(z)$) produced by our best-fit model LF model (black solid line) under the assumption of a constant $\log_{10}\xi_\mathrm{ion}\approx 25.4$ and constant $f_\mathrm{esc}=0.2$, as compared to $\dot{N}_\mathrm{ion}(z)$) models developed by Bouwens et al. \citep{Bouwens15b}, Finkelstein et al. \citep{Finkelstein19} and Naidu et al. \citep{Naidu20} (their model I) to fit a number of observational constraints on cosmic reionization. Because our best-fit LF model implies fewer low-luminosity galaxies at the highest redshifts compared to most other LFs in the literature, our ionizing emissivity is lower at $z>9$ than the other models shown here, yet is still able to complete cosmic reionization by $z\approx 6.5$.   
{\bf Right:} The cosmic neutral hydrogen fraction ($1-Q_\mathrm{HII}$) as a function of redshift for the same constant-$\xi_\mathrm{ion}$, constant-$f_\mathrm{esc}$ model (black solid line) as in the left panel. As seen the model is in rough agreement with a number of constraints on the neutral fraction, including high-redshift quasars (green symbols), the Ly$\alpha$ emitter equivalent width distribution (blue symbols for measurements and arrow for lower limit), Ly$\alpha$ emitter clustering (maroon arrow for upper limit) and the Ly$\alpha$, Ly$\beta$ dark fraction (yellow arrow for upper limit).}
\label{reionization_fig}
\end{figure*}

Our best-fitting peak mean SFEs are a factor $\sim 1-2$ times some other recent determinations \citep{SunFurlanetto16,Furlanetto2017,2019ApJ...878..114R}, but drops faster toward the mass tails. However, taking statistical uncertainty into account, we are in good agreement with these results. 

The best-fitting galaxy LF at $z = 6$ (see Fig.~\ref{fig:lumfunc}) follows closely the results in \cite{Naidu20}. The ionizing emissivity, hydrogen neutral fraction and reionization optical depth resulting from our galaxy LFs can also be made to match theirs well (although the ionizing emissivity is lower than theirs at the highest redshifts -- see Sect.~\ref{sec:reion}). They assume, like in this work, a redshift-independent SFE, but also include in their model the effects of merger histories and individual spectral energy distributions of galaxies. The close similarity of these results suggests that the impact on the galaxy LF of those physical effects can indeed be effectively described by the model we use: a magnitude scatter prescription around a mean magnitude - halo mass relation.

\subsection{Implications for reionization} \label{sec:reion}
While our fitting procedure ensures a good fit of the galaxy LF (Eq.~\ref{eq:phi}) to the luminosity function constraints derived from galaxy number counts \citep{2015ApJ...803...34B} inside the brightness and redshift range covered by the observations, there is no guarantee that the extrapolation of this LF to fainter magnitudes or higher redshifts will produce a galaxy population capable of reionizing the Universe in accordance with current constraints.

The galaxy contribution to the ionizing photon budget of the Universe is regulated by:
\begin{equation}
\dot{N}_\mathrm{ion}(z)=\rho_{\rm UV}(z)\xi_\mathrm{ion}f_\mathrm{esc},
\label{eq:Nion}
\end{equation}
where $N_\mathrm{ion}$ is the ionizing emissivity (ionizing photons per time and comoving volume), $\rho_{\rm UV}(z)$ is the dust-corrected, non-ionizing (1500 \AA) luminosity density produced by the whole galaxy population at redshift $z$, $\xi_\mathrm{ion}$ is the ionizing photon production efficiency (a conversion factor between the non-ionizing UV flux at 1500 \AA{} and the number of hydrogen-ionizing (Lyman continuum) photons),  and $f_\mathrm{esc}$ is a the fraction of ionizing photons able to evade absorption of gas and dust within galaxies and make it into the intergalactic medium. For simplicity, we have here assumed $\xi_\mathrm{ion}$ and $f_\mathrm{esc}$ to be independent of redshift and galaxy luminosity \citep[for more advanced models, see e.g.,][]{Finkelstein19,Naidu20,Yung20}.  The comoving luminosity density $\rho_\mathrm{UV}(z)$ can be derived by integrating over the dust-corrected LF at each redshift, usually with a cut-off at some faint luminosity limit where galaxy formation is assumed to become inefficient, either because of the HI cooling limit of dark matter halos or by photo-suppression by ultraviolet radiation field built up during the cosmic reionization process itself. Here, we have adopted the commonly adopted limit $M_{\rm UV}\approx -13$, and tested that going faintward has no significant impact on our result.

The $\xi_\mathrm{ion}$ parameter depends on the spectral energy distribution of galaxies, but is typically assumed to be in the range $\log_{10}\xi_\mathrm{ion}\approx 25.2$--25.8, whereas the Lyman continuum escape fraction is usually assumed to be $f_\mathrm{esc}\approx 0.01$--0.3 at $z=5$--15. 

Our machinery gives rise to a galaxy LF with a smaller number of low-luminosity galaxies at the highest redshifts (below the detection limit) compared to most LF fits in the literature, and because of this, the product $\xi_\mathrm{ion} f_\mathrm{esc}$ in Eq.~(\ref{eq:Nion}) needs to be set high to produce a hydrogen reionization history in agreement with current constraints. In the left panel of Fig.~\ref{reionization_fig}, we show the $\dot{N}_\mathrm{ion}(z)$ evolution predicted under the assumption of $\xi_\mathrm{ion}=25.4$ and $f_\mathrm{esc}=0.2$. This is contrasted to three recent models \citep{Bouwens15b,Finkelstein19,Naidu20} designed to fit a number of reionization constraints. With our parameter choices, the model a similar ionizing emissivity as these models at $z\approx 8$, but a higher $\dot{N}_\mathrm{ion}$ at lower redshifts and a much lower one at higher redshifts. As is to be expected, our emissivity model most closely resembles that of \cite{Naidu20}, where high-luminosity galaxies are similarly invoked to provide most of the ionizing photons affecting the IGM.

To derive the evolution of the cosmic ionized volume fraction $Q_\mathrm{HII}$ we solve the differential equation 
\begin{equation}
\dot{Q}_\mathrm{HII}(z)=\frac{\dot{N}_\mathrm{ion}(z)}{\overline{n}_\mathrm{H}} - \frac{Q_\mathrm{HII}}{t_\mathrm{rec}},
\label{eq:Q}
\end{equation}
where $\overline{n}_\mathrm{H}$ is the comoving cosmic hydrogen density (here assumed to be $ \approx 1.9\times 10^{-7}$ cm$^{-3}$; \citet{2014ARA&A..52..415M}) and $t_\mathrm{rec}$ is the IGM recombination time of hydrogen (computed as in \citet{2014ARA&A..52..415M}, but with redshift-dependent clumping factors at $z=6$--14 based on the LN25N512 simulation of \citet{Pawlik15}). This results in the redshift evolution of the cosmic neutral hydrogen fraction (1-$Q_\mathrm{HII}$) shown in the right panel of 
Fig.~\ref{reionization_fig}. This reionization history turns out to be in rough agreement with observational constraints based on the dark pixel fraction of the Ly$\alpha$, L$\beta$ forest \citep{McGreer15}, the equivalent width distribution of Ly$\alpha$-emitters \citep{Mason18,Hoag19,Mason19}, the clustering of Ly$\alpha$-emitters \citep{Ouchi10} and individual high-redshift quasar sightlines \citep{Davies18,Yang20, Wang20}. Our model also  produces a Thomson scattering optical depth of $\tau \approx 0.048$ in agreement with the Planck+18 measurement of $\tau=0.054\pm0.007$ \citep{Planck18}, and a midpoint of reionization (1-$Q_\mathrm{HII}$=0.5) at $z\approx 7.1$. 

Even though this simple model -- primarily based on our best-fit LF derived completely without reionziation considerations -- manages to achieve hydrogen reionization by $z\approx 6.5$ and produce evolution in the cosmic neutral fraction in rough agreement with existing constraints, it overshoots the $z\approx5$ ionizing emissivity measurement of \cite{Becker13} by a factor of $\approx 4$, which means that  $\xi_\mathrm{ion} f_\mathrm{esc}$ would have to drop by a similar factor in no more than $\approx 350$ Myr once reionization is complete. While this may not be completely implausible, we do not currently have any physical justification for such abrupt parameter evolution at these redshifts. 
Hence, taking hydrogen reionization constraints into account during the fitting procedure (instead of being used as a posteriori consistency check) would likely alter the best-fitting LF by boosting the comoving number density of low-luminosity galaxies at the high-redshift end \cite[e.g.,][]{Park2019, 2020MNRAS.491.3891P}. The SFE model assumptions may also not be fully realistic. For example, the SFE at low halo masses (that we are here not observationally sensitive to) may deviate from the assumed power-law shape. Also, the UV magnitude scatter is realistically not constant in both mass and redshift. An increased scatter at high redshift could help boost the ionizing emissivity. 


\section{SUMMARY}
We have introduced a new semi-analytical framework for modelling early-Universe galaxy population statistics that combines sensitivity to cosmological and star formation models. The framework is implemented in the novel parallelized code GalaxyMC, and publicly available in the cloud computing application \url{galaxymc.cloud}. We use GalaxyMC to derive MCMC parameter constraints on an 11-parameter model of star formation, cosmology and systematics with galaxy UV luminosity and correlation function data at $z = 4 - 10$. 

For the first time using data at these redshifts, we constrain 
\begin{align}
\sigma_8 = 0.81 \pm 0.03\,, & & & \,\,{\rm I.\,H_0\,low-}z \\
\sigma_8 = 0.84 \pm 0.03\,, & & & \,\,{\rm II.\,H_0\, Planck}  \\
\sigma_8 = 0.81 \pm 0.06\,, & & & \,\,{\rm IV.}\,h\leq 1
\end{align}
with a BBN prior on $\Omega_{\rm b}h^2$ and a SNIa prior on $\Omega_{\rm m}$. Without a prior on $\Omega_{\rm m}$, we find
\begin{align} 
\sigma_8 = 0.78 \pm 0.06\,, & & & \,\,{\rm III.\,H_0\,low-}z \\
\Omega_{\rm m} = 0.33 \pm 0.07\,. & & &  
\end{align}
The inferred parameter values are consistent with constraints from other contemporary cosmological data, albeit with marginal tension between the {\it Planck} Hubble constant prior case (II) and some late-time Universe measurements of $\sigma_8$.  The preferred values of the freely fitted $h$ (IV) are closer to those of the low-$z$ Hubble constant prior than the {\it Planck} prior, though consistent with both. In summary, we find no evidence at $z = 4$--$10$ for a deviation from low-redshift cosmology, but possible hints of tension with a {\it Planck} 2018 cosmology.

With a low-$z$ or {\it Planck} Hubble constant prior (I, II), the peak mean SFE is constrained to be $\sim (19 \pm 5)\%$ at $z \sim 10$, rising to $\sim (32 \pm 4)\%$ at $z \sim 4$, given the double power-law SFE model assumed, and consistent with other theoretical and observational estimates a UV magnitude scatter of $\sigma_{M} = 0.56 \pm 0.08$ or $\sigma_{M} = 0.48 \pm 0.09$. The constraints on the five SFE model parameters are quite similar and well-constrained for all the data/prior combinations, except for the high-mass slope $\alpha$ and UV magnitude scatter $\sigma_{M}$ in the case with no prior on $h$ (IV), due to significant degeneracies. Likewise, the mean SFEs and UV magnitudes as a function of halo mass show essentially no difference between a low-$z$ or {\it Planck} Hubble constant prior (I, II), and are in good agreement with other recent estimates in the literature. 

Our best-fit model predicts fewer galaxies at the faint end of the LF at $z\gtrsim 7$ compared to the reference LFs of \citet{Finkelstein16}, but can still achieve a Thomson scattering optical depth to reionization of $\tau = 0.048$, in good agreement with {\it Planck} 2018, under reasonable assumptions on the escape fraction of ionizing photons and the ionizing photon production efficiency. The corresponding ionizing emissivity exhibits stronger redshift evolution than those of \cite{Bouwens15b} and \cite{Finkelstein19}, but is similar to that of \cite{Naidu20}, and reaches a mid-point of reionization at $z\approx 7.1$. 

This study presents the first simultaneous constraints on cosmology and star formation in the $z>4$ post-CMB Universe. This epoch in cosmic history is a relatively unexplored bridge between late-Universe (e.g. galaxy surveys, SNIa) and early-Universe cosmological probes (e.g. BBN, CMB). Studying galaxy population statistics during this epoch particularly can provide new information about the matter distribution on small scales, and the expansion and growth history at high redshift. The advent of deeper and larger surveys of this epoch with the {\it James Webb Space Telescope}, {\it EUCLID}, and later the {\it Roman Space Telescope} and the Extremely Large Telescope (ELT), will enable tests of new physics in the $z>4$ Universe that could also shed light on the current observational tension in measurements of the Hubble constant $h$ and matter power spectrum normalization $\sigma_8$.

In ongoing and future work, we plan to employ the full range of available high-redshift galaxy and 21cm data in synergy to constrain cosmology and star formation, and test models of new physics in the early Universe. 

\vspace{5mm}
{\bf Acknowledgments}
We thank P.~Behroozi, R.~Bouwens, L.~Bradley, D.~Coe, J.~Dunlop, E.~Macaulay, G.~Mamon, C.~Maraston, A.~Mesinger, L.~Moustakas, J.~Silk, D.~Spolyar and L.Y.A.~Yung for helpful conversations related to this work. 

The computations were enabled by resources provided by the Swedish National Infrastructure for Computing (SNIC) at Uppsala Multidisciplinary Center for Advanced Computational Science (UPPMAX) under projects SNIC 2018/3-667 and SNIC 2020/15-25, partially funded by the Swedish Research Council through grant agreement no. 2016-07213. The computations and availability of GalaxyMC Cloud are enabled by resources provided by SNIC at Chalmers Centre for Computational Science and Engineering (C3SE), High Performance Computing Center North (HPC2N) and UPPMAX under projects SNIC 2020/20-2 and SNIC 2021/18-10, partially funded by the Swedish Research Council through grant agreements no. 2016-07213 and 2018-05973. Marcus Lundberg at UPPMAX is acknowledged for assistance concerning technical and implementational aspects in making the code run on the UPPMAX resources. Mathias Lindberg and Lars Viklund at SNIC Science Cloud are acknowledged for assistance concerning technical and implementational aspects in making the code run on the SNIC Science Cloud resources.

MS was supported by the Fulbright Commission, Helge Ax:son Johnson Foundation, the Lars Hierta Memorial Fund, the Lundstr\"om--\r{A}man Foundation, the L\"angmanska Fund for Culture, and the Olle Engkvist Foundation. MS also acknowledges the award of a P.E.~Fil\'en fellowship from the Fil\'en Foundation at Uppsala University, and a Natural Sciences fellowship at the Swedish Collegium for Advanced Study.

EZ acknowledges funding from the Swedish National Space Board.

\vspace{5mm}
\software{CAMB \citep{2000ApJ...538..473L}, Cuba \citep{2005CoPhC.168...78H, HAHN2016341}, NumPy \citep{2020Natur.585..357H}, SciPy \citep{2020NatMe..17..261V}, Matplotlib \citep{2007CSE.....9...90H}, emcee \citep{2019JOSS....4.1864F}, corner \citep{corner}}


\bibliographystyle{aasjournal}

\begin{thebibliography}{}
\expandafter\ifx\csname natexlab\endcsname\relax\def\natexlab#1{#1}\fi
\providecommand{\url}[1]{\href{#1}{#1}}
\providecommand{\dodoi}[1]{doi:~\href{http://doi.org/#1}{\nolinkurl{#1}}}
\providecommand{\doeprint}[1]{\href{http://ascl.net/#1}{\nolinkurl{http://ascl.net/#1}}}
\providecommand{\doarXiv}[1]{\href{https://arxiv.org/abs/#1}{\nolinkurl{https://arxiv.org/abs/#1}}}

\bibitem[{{Becker} \& {Bolton}(2013)}]{Becker13}
{Becker}, G.~D., \& {Bolton}, J.~S. 2013, \mnras, 436, 1023,
  \dodoi{10.1093/mnras/stt1610}

\bibitem[{{Behroozi} {et~al.}(2019){Behroozi}, {Wechsler}, {Hearin}, \&
  {Conroy}}]{Behroozi2019}
{Behroozi}, P., {Wechsler}, R.~H., {Hearin}, A.~P., \& {Conroy}, C. 2019,
  \mnras, 488, 3143, \dodoi{10.1093/mnras/stz1182}

\bibitem[{Behroozi {et~al.}(2013)Behroozi, Wechsler, \& Conroy}]{Behroozi2013}
Behroozi, P.~S., Wechsler, R.~H., \& Conroy, C. 2013, \apj, 770, 57

\bibitem[{{Bhowmick} {et~al.}(2018){Bhowmick}, {Campbell}, {Di Matteo}, \&
  {Feng}}]{2018MNRAS.480.3177B}
{Bhowmick}, A.~K., {Campbell}, D., {Di Matteo}, T., \& {Feng}, Y. 2018, \mnras,
  480, 3177, \dodoi{10.1093/mnras/sty2128}

\bibitem[{{Bouwens} {et~al.}(2015{\natexlab{a}}){Bouwens}, {Illingworth},
  {Oesch}, {Caruana}, {Holwerda}, {Smit}, \& {Wilkins}}]{Bouwens15b}
{Bouwens}, R.~J., {Illingworth}, G.~D., {Oesch}, P.~A., {et~al.}
  2015{\natexlab{a}}, \apj, 811, 140, \dodoi{10.1088/0004-637X/811/2/140}

\bibitem[{{Bouwens} {et~al.}(2015{\natexlab{b}}){Bouwens}, {Illingworth},
  {Oesch}, {Trenti}, {Labb{\'e}}, {Bradley}, {Carollo}, {van Dokkum},
  {Gonzalez}, {Holwerda}, {Franx}, {Spitler}, {Smit}, \&
  {Magee}}]{2015ApJ...803...34B}
---. 2015{\natexlab{b}}, \apj, 803, 34, \dodoi{10.1088/0004-637X/803/1/34}

\bibitem[{{Bowler} {et~al.}(2020){Bowler}, {Jarvis}, {Dunlop}, {McLure},
  {McLeod}, {Adams}, {Milvang-Jensen}, \& {McCracken}}]{Bowler2020}
{Bowler}, R.~A.~A., {Jarvis}, M.~J., {Dunlop}, J.~S., {et~al.} 2020, \mnras,
  493, 2059, \dodoi{10.1093/mnras/staa313}

\bibitem[{{Corasaniti} {et~al.}(2017){Corasaniti}, {Agarwal}, {Marsh}, \&
  {Das}}]{2017PhRvD..95h3512C}
{Corasaniti}, P.~S., {Agarwal}, S., {Marsh}, D.~J.~E., \& {Das}, S. 2017, \prd,
  95, 083512, \dodoi{10.1103/PhysRevD.95.083512}

\bibitem[{{Correa} {et~al.}(2015){Correa}, {Wyithe}, {Schaye}, \&
  {Duffy}}]{2015MNRAS.452.1217C}
{Correa}, C.~A., {Wyithe}, J. S.~B., {Schaye}, J., \& {Duffy}, A.~R. 2015,
  \mnras, 452, 1217, \dodoi{10.1093/mnras/stv1363}

\bibitem[{{Davies} {et~al.}(2018){Davies}, {Hennawi}, {Ba{\~n}ados},
  {Luki{\'c}}, {Decarli}, {Fan}, {Farina}, {Mazzucchelli}, {Rix}, {Venemans},
  {Walter}, {Wang}, \& {Yang}}]{Davies18}
{Davies}, F.~B., {Hennawi}, J.~F., {Ba{\~n}ados}, E., {et~al.} 2018, \apj, 864,
  142, \dodoi{10.3847/1538-4357/aad6dc}

\bibitem[{Dayal {et~al.}(2013)Dayal, Dunlop, Maio, \& Ciardi}]{Dayal2013}
Dayal, P., Dunlop, J.~S., Maio, U., \& Ciardi, B. 2013, \mnras, 434, 1486

\bibitem[{{Dayal} \& {Ferrara}(2018)}]{2018PhR...780....1D}
{Dayal}, P., \& {Ferrara}, A. 2018, \physrep, 780, 1,
  \dodoi{10.1016/j.physrep.2018.10.002}

\bibitem[{{Di Valentino} {et~al.}(2021){Di Valentino}, {Mena}, {Pan},
  {Visinelli}, {Yang}, {Melchiorri}, {Mota}, {Riess}, \&
  {Silk}}]{2021arXiv210301183D}
{Di Valentino}, E., {Mena}, O., {Pan}, S., {et~al.} 2021, arXiv e-prints,
  arXiv:2103.01183.
\newblock \doarXiv{2103.01183}

\bibitem[{{Finkelstein}(2016)}]{Finkelstein16}
{Finkelstein}, S.~L. 2016, \pasa, 33, e037, \dodoi{10.1017/pasa.2016.26}

\bibitem[{{Finkelstein} {et~al.}(2015){Finkelstein}, {Ryan}, {Papovich},
  {Dickinson}, {Song}, {Somerville}, {Ferguson}, {Salmon}, {Giavalisco},
  {Koekemoer}, {Ashby}, {Behroozi}, {Castellano}, {Dunlop}, {Faber}, {Fazio},
  {Fontana}, {Grogin}, {Hathi}, {Jaacks}, {Kocevski}, {Livermore}, {McLure},
  {Merlin}, {Mobasher}, {Newman}, {Rafelski}, {Tilvi}, \&
  {Willner}}]{Finkelstein2015}
{Finkelstein}, S.~L., {Ryan}, Jr., R.~E., {Papovich}, C., {et~al.} 2015, \apj,
  810, 71, \dodoi{10.1088/0004-637X/810/1/71}

\bibitem[{{Finkelstein} {et~al.}(2019){Finkelstein}, {D'Aloisio},
  {Paardekooper}, {Ryan}, {Behroozi}, {Finlator}, {Livermore}, {Upton
  Sanderbeck}, {Dalla Vecchia}, \& {Khochfar}}]{Finkelstein19}
{Finkelstein}, S.~L., {D'Aloisio}, A., {Paardekooper}, J.-P., {et~al.} 2019,
  \apj, 879, 36, \dodoi{10.3847/1538-4357/ab1ea8}

\bibitem[{Foreman-Mackey(2016)}]{corner}
Foreman-Mackey, D. 2016, The Journal of Open Source Software, 1, 24,
  \dodoi{10.21105/joss.00024}

\bibitem[{{Foreman-Mackey} {et~al.}(2013){Foreman-Mackey}, {Hogg}, {Lang}, \&
  {Goodman}}]{2013PASP..125..306F}
{Foreman-Mackey}, D., {Hogg}, D.~W., {Lang}, D., \& {Goodman}, J. 2013, \pasp,
  125, 306, \dodoi{10.1086/670067}

\bibitem[{{Foreman-Mackey} {et~al.}(2019){Foreman-Mackey}, {Farr}, {Sinha},
  {Archibald}, {Hogg}, {Sanders}, {Zuntz}, {Williams}, {Nelson}, {de
  Val-Borro}, {Erhardt}, {Pashchenko}, \& {Pla}}]{2019JOSS....4.1864F}
{Foreman-Mackey}, D., {Farr}, W., {Sinha}, M., {et~al.} 2019, The Journal of
  Open Source Software, 4, 1864, \dodoi{10.21105/joss.01864}

\bibitem[{{Furlanetto} {et~al.}(2017){Furlanetto}, {Mirocha}, {Mebane}, \&
  {Sun}}]{Furlanetto2017}
{Furlanetto}, S.~R., {Mirocha}, J., {Mebane}, R.~H., \& {Sun}, G. 2017, \mnras,
  472, 1576, \dodoi{10.1093/mnras/stx2132}

\bibitem[{{Goodman} \& {Weare}(2010)}]{2010CAMCS...5...65G}
{Goodman}, J., \& {Weare}, J. 2010, Communications in Applied Mathematics and
  Computational Science, 5, 65, \dodoi{10.2140/camcos.2010.5.65}

\bibitem[{{Hahn}(2005)}]{2005CoPhC.168...78H}
{Hahn}, T. 2005, Computer Physics Communications, 168, 78,
  \dodoi{10.1016/j.cpc.2005.01.010}

\bibitem[{Hahn(2016)}]{HAHN2016341}
Hahn, T. 2016, Computer Physics Communications, 207, 341 ,
  \dodoi{https://doi.org/10.1016/j.cpc.2016.05.012}

\bibitem[{{Harikane} {et~al.}(2018){Harikane}, {Ouchi}, {Ono}, {Saito},
  {Behroozi}, {More}, {Shimasaku}, {Toshikawa}, {Lin}, {Akiyama}, {Coupon},
  {Komiyama}, {Konno}, {Lin}, {Miyazaki}, {Nishizawa}, {Shibuya}, \&
  {Silverman}}]{Harikane2018}
{Harikane}, Y., {Ouchi}, M., {Ono}, Y., {et~al.} 2018, \pasj, 70, S11,
  \dodoi{10.1093/pasj/psx097}

\bibitem[{{Harris} {et~al.}(2020){Harris}, {Millman}, {van der Walt},
  {Gommers}, {Virtanen}, {Cournapeau}, {Wieser}, {Taylor}, {Berg}, {Smith},
  {Kern}, {Picus}, {Hoyer}, {van Kerkwijk}, {Brett}, {Haldane}, {del R{\'\i}o},
  {Wiebe}, {Peterson}, {G{\'e}rard-Marchant}, {Sheppard}, {Reddy}, {Weckesser},
  {Abbasi}, {Gohlke}, \& {Oliphant}}]{2020Natur.585..357H}
{Harris}, C.~R., {Millman}, K.~J., {van der Walt}, S.~J., {et~al.} 2020, \nat,
  585, 357, \dodoi{10.1038/s41586-020-2649-2}

\bibitem[{{Hatfield} {et~al.}(2018){Hatfield}, {Bowler}, {Jarvis}, \&
  {Hale}}]{Hatfield2018}
{Hatfield}, P.~W., {Bowler}, R.~A.~A., {Jarvis}, M.~J., \& {Hale}, C.~L. 2018,
  \mnras, 477, 3760, \dodoi{10.1093/mnras/sty856}

\bibitem[{{Heymans} {et~al.}(2021){Heymans}, {Tr{\"o}ster}, {Asgari}, {Blake},
  {Hildebrandt}, {Joachimi}, {Kuijken}, {Lin}, {S{\'a}nchez}, {van den Busch},
  {Wright}, {Amon}, {Bilicki}, {de Jong}, {Crocce}, {Dvornik}, {Erben},
  {Fortuna}, {Getman}, {Giblin}, {Glazebrook}, {Hoekstra}, {Joudaki},
  {Kannawadi}, {K{\"o}hlinger}, {Lidman}, {Miller}, {Napolitano}, {Parkinson},
  {Schneider}, {Shan}, {Valentijn}, {Verdoes Kleijn}, \&
  {Wolf}}]{2021A&A...646A.140H}
{Heymans}, C., {Tr{\"o}ster}, T., {Asgari}, M., {et~al.} 2021, \aap, 646, A140,
  \dodoi{10.1051/0004-6361/202039063}

\bibitem[{{Hoag} {et~al.}(2019){Hoag}, {Brada{\v{c}}}, {Huang}, {Mason},
  {Treu}, {Schmidt}, {Trenti}, {Strait}, {Lemaux}, {Finney}, \&
  {Paddock}}]{Hoag19}
{Hoag}, A., {Brada{\v{c}}}, M., {Huang}, K., {et~al.} 2019, \apj, 878, 12,
  \dodoi{10.3847/1538-4357/ab1de7}

\bibitem[{{Hunter}(2007)}]{2007CSE.....9...90H}
{Hunter}, J.~D. 2007, Computing in Science and Engineering, 9, 90,
  \dodoi{10.1109/MCSE.2007.55}

\bibitem[{{Ishigaki} {et~al.}(2018){Ishigaki}, {Kawamata}, {Ouchi}, {Oguri},
  {Shimasaku}, \& {Ono}}]{Ishigaki18}
{Ishigaki}, M., {Kawamata}, R., {Ouchi}, M., {et~al.} 2018, \apj, 854, 73,
  \dodoi{10.3847/1538-4357/aaa544}

\bibitem[{{Krauss} \& {Chaboyer}(2003)}]{2003Sci...299...65K}
{Krauss}, L.~M., \& {Chaboyer}, B. 2003, Science, 299, 65,
  \dodoi{10.1126/science.1075631}

\bibitem[{{Lewis} {et~al.}(2000){Lewis}, {Challinor}, \&
  {Lasenby}}]{2000ApJ...538..473L}
{Lewis}, A., {Challinor}, A., \& {Lasenby}, A. 2000, \apj, 538, 473,
  \dodoi{10.1086/309179}

\bibitem[{Livermore {et~al.}(2018)Livermore, Trenti, Bradley, Bernard,
  Holwerda, Mason, \& Treu}]{Livermore2018}
Livermore, R.~C., Trenti, M., Bradley, L.~D., {et~al.} 2018, \apjl, 861, L17

\bibitem[{{Madau} \& {Dickinson}(2014)}]{2014ARA&A..52..415M}
{Madau}, P., \& {Dickinson}, M. 2014, \araa, 52, 415,
  \dodoi{10.1146/annurev-astro-081811-125615}

\bibitem[{{Mason} {et~al.}(2015){Mason}, {Trenti}, \& {Treu}}]{Mason2015}
{Mason}, C.~A., {Trenti}, M., \& {Treu}, T. 2015, \apj, 813, 21,
  \dodoi{10.1088/0004-637X/813/1/21}

\bibitem[{{Mason} {et~al.}(2018){Mason}, {Treu}, {Dijkstra}, {Mesinger},
  {Trenti}, {Pentericci}, {de Barros}, \& {Vanzella}}]{Mason18}
{Mason}, C.~A., {Treu}, T., {Dijkstra}, M., {et~al.} 2018, \apj, 856, 2,
  \dodoi{10.3847/1538-4357/aab0a7}

\bibitem[{{Mason} {et~al.}(2019){Mason}, {Fontana}, {Treu}, {Schmidt}, {Hoag},
  {Abramson}, {Amorin}, {Brada{\v{c}}}, {Guaita}, {Jones}, {Henry}, {Malkan},
  {Pentericci}, {Trenti}, \& {Vanzella}}]{Mason19}
{Mason}, C.~A., {Fontana}, A., {Treu}, T., {et~al.} 2019, \mnras, 485, 3947,
  \dodoi{10.1093/mnras/stz632}

\bibitem[{{McGreer} {et~al.}(2015){McGreer}, {Mesinger}, \&
  {D'Odorico}}]{McGreer15}
{McGreer}, I.~D., {Mesinger}, A., \& {D'Odorico}, V. 2015, \mnras, 447, 499,
  \dodoi{10.1093/mnras/stu2449}

\bibitem[{McLeod {et~al.}(2016)McLeod, McLure, \& Dunlop}]{McLeod2016}
McLeod, D.~J., McLure, R.~J., \& Dunlop, J.~S. 2016, \mnras, 459, 3812

\bibitem[{{Menci} {et~al.}(2016){Menci}, {Grazian}, {Castellano}, \&
  {Sanchez}}]{2016ApJ...825L...1M}
{Menci}, N., {Grazian}, A., {Castellano}, M., \& {Sanchez}, N.~G. 2016, \apjl,
  825, L1, \dodoi{10.3847/2041-8205/825/1/L1}

\bibitem[{{Menci} {et~al.}(2017){Menci}, {Merle}, {Totzauer}, {Schneider},
  {Grazian}, {Castellano}, \& {Sanchez}}]{2017ApJ...836...61M}
{Menci}, N., {Merle}, A., {Totzauer}, M., {et~al.} 2017, \apj, 836, 61,
  \dodoi{10.3847/1538-4357/836/1/61}

\bibitem[{{Menci} {et~al.}(2020){Menci}, {Grazian}, {Castellano}, {Santini},
  {Giallongo}, {Lamastra}, {Fortuni}, {Fontana}, {Merlin}, {Wang}, {Elbaz}, \&
  {Sanchez}}]{2020ApJ...900..108M}
{Menci}, N., {Grazian}, A., {Castellano}, M., {et~al.} 2020, \apj, 900, 108,
  \dodoi{10.3847/1538-4357/aba9d2}

\bibitem[{{Mirocha}(2020)}]{Mirocha2020b}
{Mirocha}, J. 2020, \mnras, 499, 4534, \dodoi{10.1093/mnras/staa3150}

\bibitem[{{Mirocha} {et~al.}(2020){Mirocha}, {Mason}, \&
  {Stark}}]{2020MNRAS.498.2645M}
{Mirocha}, J., {Mason}, C., \& {Stark}, D.~P. 2020, \mnras, 498, 2645,
  \dodoi{10.1093/mnras/staa2586}

\bibitem[{Morishita {et~al.}(2018)Morishita, Trenti, Stiavelli, Bradley, Coe,
  Oesch, Mason, Bridge, Holwerda, Livermore, Salmon, Schmidt, Shull, \&
  Treu}]{Morishita2018}
Morishita, T., Trenti, M., Stiavelli, M., {et~al.} 2018, \apj, 867, 150

\bibitem[{Moster {et~al.}(2010)Moster, Somerville, Maulbetsch, van~den Bosch,
  Macci{\`o}, Naab, \& Oser}]{Moster2010}
Moster, B.~P., Somerville, R.~S., Maulbetsch, C., {et~al.} 2010, \apj, 710, 903

\bibitem[{{Naidu} {et~al.}(2020){Naidu}, {Tacchella}, {Mason}, {Bose}, {Oesch},
  \& {Conroy}}]{Naidu20}
{Naidu}, R.~P., {Tacchella}, S., {Mason}, C.~A., {et~al.} 2020, \apj, 892, 109,
  \dodoi{10.3847/1538-4357/ab7cc9}

\bibitem[{{Oesch} {et~al.}(2018){Oesch}, {Bouwens}, {Illingworth}, {Labb{\'e}},
  \& {Stefanon}}]{Oesch2018}
{Oesch}, P.~A., {Bouwens}, R.~J., {Illingworth}, G.~D., {Labb{\'e}}, I., \&
  {Stefanon}, M. 2018, \apj, 855, 105, \dodoi{10.3847/1538-4357/aab03f}

\bibitem[{{Oke}(1974)}]{1974ApJS...27...21O}
{Oke}, J.~B. 1974, \apjs, 27, 21, \dodoi{10.1086/190287}

\bibitem[{{Ouchi} {et~al.}(2010){Ouchi}, {Shimasaku}, {Furusawa}, {Saito},
  {Yoshida}, {Akiyama}, {Ono}, {Yamada}, {Ota}, {Kashikawa}, {Iye}, {Kodama},
  {Okamura}, {Simpson}, \& {Yoshida}}]{Ouchi10}
{Ouchi}, M., {Shimasaku}, K., {Furusawa}, H., {et~al.} 2010, \apj, 723, 869,
  \dodoi{10.1088/0004-637X/723/1/869}

\bibitem[{{Pacucci} {et~al.}(2013){Pacucci}, {Mesinger}, \&
  {Haiman}}]{2013MNRAS.435L..53P}
{Pacucci}, F., {Mesinger}, A., \& {Haiman}, Z. 2013, \mnras, 435, L53,
  \dodoi{10.1093/mnrasl/slt093}

\bibitem[{{Park} {et~al.}(2020){Park}, {Gillet}, {Mesinger}, \&
  {Greig}}]{2020MNRAS.491.3891P}
{Park}, J., {Gillet}, N., {Mesinger}, A., \& {Greig}, B. 2020, \mnras, 491,
  3891, \dodoi{10.1093/mnras/stz3278}

\bibitem[{{Park} {et~al.}(2019){Park}, {Mesinger}, {Greig}, \&
  {Gillet}}]{Park2019}
{Park}, J., {Mesinger}, A., {Greig}, B., \& {Gillet}, N. 2019, \mnras, 484,
  933, \dodoi{10.1093/mnras/stz032}

\bibitem[{{Pawlik} {et~al.}(2015){Pawlik}, {Schaye}, \& {Dalla
  Vecchia}}]{Pawlik15}
{Pawlik}, A.~H., {Schaye}, J., \& {Dalla Vecchia}, C. 2015, \mnras, 451, 1586,
  \dodoi{10.1093/mnras/stv976}

\bibitem[{{Planck Collaboration} {et~al.}(2020){Planck Collaboration},
  {Aghanim}, {Akrami}, {Ashdown}, {Aumont}, {Baccigalupi}, {Ballardini},
  {Banday}, {Barreiro}, {Bartolo}, {Basak}, {Battye}, {Benabed}, {Bernard},
  {Bersanelli}, {Bielewicz}, {Bock}, {Bond}, {Borrill}, {Bouchet}, {Boulanger},
  {Bucher}, {Burigana}, {Butler}, {Calabrese}, {Cardoso}, {Carron},
  {Challinor}, {Chiang}, {Chluba}, {Colombo}, {Combet}, {Contreras}, {Crill},
  {Cuttaia}, {de Bernardis}, {de Zotti}, {Delabrouille}, {Delouis}, {Di
  Valentino}, {Diego}, {Dor{\'e}}, {Douspis}, {Ducout}, {Dupac}, {Dusini},
  {Efstathiou}, {Elsner}, {En{\ss}lin}, {Eriksen}, {Fantaye}, {Farhang},
  {Fergusson}, {Fernandez-Cobos}, {Finelli}, {Forastieri}, {Frailis},
  {Fraisse}, {Franceschi}, {Frolov}, {Galeotta}, {Galli}, {Ganga},
  {G{\'e}nova-Santos}, {Gerbino}, {Ghosh}, {Gonz{\'a}lez-Nuevo}, {G{\'o}rski},
  {Gratton}, {Gruppuso}, {Gudmundsson}, {Hamann}, {Handley}, {Hansen},
  {Herranz}, {Hildebrandt}, {Hivon}, {Huang}, {Jaffe}, {Jones}, {Karakci},
  {Keih{\"a}nen}, {Keskitalo}, {Kiiveri}, {Kim}, {Kisner}, {Knox},
  {Krachmalnicoff}, {Kunz}, {Kurki-Suonio}, {Lagache}, {Lamarre}, {Lasenby},
  {Lattanzi}, {Lawrence}, {Le Jeune}, {Lemos}, {Lesgourgues}, {Levrier},
  {Lewis}, {Liguori}, {Lilje}, {Lilley}, {Lindholm}, {L{\'o}pez-Caniego},
  {Lubin}, {Ma}, {Mac{\'\i}as-P{\'e}rez}, {Maggio}, {Maino}, {Mandolesi},
  {Mangilli}, {Marcos-Caballero}, {Maris}, {Martin}, {Martinelli},
  {Mart{\'\i}nez-Gonz{\'a}lez}, {Matarrese}, {Mauri}, {McEwen}, {Meinhold},
  {Melchiorri}, {Mennella}, {Migliaccio}, {Millea}, {Mitra},
  {Miville-Desch{\^e}nes}, {Molinari}, {Montier}, {Morgante}, {Moss}, {Natoli},
  {N{\o}rgaard-Nielsen}, {Pagano}, {Paoletti}, {Partridge}, {Patanchon},
  {Peiris}, {Perrotta}, {Pettorino}, {Piacentini}, {Polastri}, {Polenta},
  {Puget}, {Rachen}, {Reinecke}, {Remazeilles}, {Renzi}, {Rocha}, {Rosset},
  {Roudier}, {Rubi{\~n}o-Mart{\'\i}n}, {Ruiz-Granados}, {Salvati}, {Sandri},
  {Savelainen}, {Scott}, {Shellard}, {Sirignano}, {Sirri}, {Spencer},
  {Sunyaev}, {Suur-Uski}, {Tauber}, {Tavagnacco}, {Tenti}, {Toffolatti},
  {Tomasi}, {Trombetti}, {Valenziano}, {Valiviita}, {Van Tent}, {Vibert},
  {Vielva}, {Villa}, {Vittorio}, {Wandelt}, {Wehus}, {White}, {White},
  {Zacchei}, \& {Zonca}}]{Planck18}
{Planck Collaboration}, {Aghanim}, N., {Akrami}, Y., {et~al.} 2020, \aap, 641,
  A6, \dodoi{10.1051/0004-6361/201833910}

\bibitem[{{Ren} {et~al.}(2019){Ren}, {Trenti}, \&
  {Mason}}]{2019ApJ...878..114R}
{Ren}, K., {Trenti}, M., \& {Mason}, C.~A. 2019, \apj, 878, 114,
  \dodoi{10.3847/1538-4357/ab2117}

\bibitem[{{Ren} {et~al.}(2018){Ren}, {Trenti}, \&
  {Mutch}}]{2018ApJ...856...81R}
{Ren}, K., {Trenti}, M., \& {Mutch}, S.~J. 2018, \apj, 856, 81,
  \dodoi{10.3847/1538-4357/aab094}

\bibitem[{{Riess} {et~al.}(2018){Riess}, {Casertano}, {Yuan}, {Macri},
  {Anderson}, {MacKenty}, {Bowers}, {Clubb}, {Filippenko}, {Jones}, \&
  {Tucker}}]{2018ApJ...855..136R}
{Riess}, A.~G., {Casertano}, S., {Yuan}, W., {et~al.} 2018, \apj, 855, 136,
  \dodoi{10.3847/1538-4357/aaadb7}

\bibitem[{{Rodr{\'\i}guez-Puebla} {et~al.}(2016){Rodr{\'\i}guez-Puebla},
  {Behroozi}, {Primack}, {Klypin}, {Lee}, \& {Hellinger}}]{2016MNRAS.462..893R}
{Rodr{\'\i}guez-Puebla}, A., {Behroozi}, P., {Primack}, J., {et~al.} 2016,
  \mnras, 462, 893, \dodoi{10.1093/mnras/stw1705}

\bibitem[{{Rojas-Ruiz} {et~al.}(2020){Rojas-Ruiz}, {Finkelstein}, {Bagley},
  {Stevans}, {Finkelstein}, {Larson}, {Mechtley}, \&
  {Diekmann}}]{RojasRuiz2020}
{Rojas-Ruiz}, S., {Finkelstein}, S.~L., {Bagley}, M.~B., {et~al.} 2020, \apj,
  891, 146, \dodoi{10.3847/1538-4357/ab7659}

\bibitem[{{Sabti} {et~al.}(2021){Sabti}, {Mu{\~n}oz}, \&
  {Blas}}]{2021JCAP...01..010S}
{Sabti}, N., {Mu{\~n}oz}, J.~B., \& {Blas}, D. 2021, \jcap, 2021, 010,
  \dodoi{10.1088/1475-7516/2021/01/010}

\bibitem[{{Sahl{\'e}n} {et~al.}(2009){Sahl{\'e}n}, {Viana}, {Liddle}, \&
  {Romer}}]{SahlenEqs}
{Sahl{\'e}n}, M., {Viana}, P.~T.~P., {Liddle}, A.~R., \& {Romer}, A.~K. 2009,
  \mnras, 397, 577, \dodoi{10.1111/j.1365-2966.2009.14923.x}

\bibitem[{{Schultz} {et~al.}(2014){Schultz}, {O{\~n}orbe}, {Abazajian}, \&
  {Bullock}}]{2014MNRAS.442.1597S}
{Schultz}, C., {O{\~n}orbe}, J., {Abazajian}, K.~N., \& {Bullock}, J.~S. 2014,
  \mnras, 442, 1597, \dodoi{10.1093/mnras/stu976}

\bibitem[{{Scolnic} {et~al.}(2018){Scolnic}, {Jones}, {Rest}, {Pan},
  {Chornock}, {Foley}, {Huber}, {Kessler}, {Narayan}, {Riess}, {Rodney},
  {Berger}, {Brout}, {Challis}, {Drout}, {Finkbeiner}, {Lunnan}, {Kirshner},
  {Sand ers}, {Schlafly}, {Smartt}, {Stubbs}, {Tonry}, {Wood-Vasey}, {Foley},
  {Hand}, {Johnson}, {Burgett}, {Chambers}, {Draper}, {Hodapp}, {Kaiser},
  {Kudritzki}, {Magnier}, {Metcalfe}, {Bresolin}, {Gall}, {Kotak}, {McCrum}, \&
  {Smith}}]{2018ApJ...859..101S}
{Scolnic}, D.~M., {Jones}, D.~O., {Rest}, A., {et~al.} 2018, \apj, 859, 101,
  \dodoi{10.3847/1538-4357/aab9bb}

\bibitem[{{Sharma}(2017)}]{2017ARA&A..55..213S}
{Sharma}, S. 2017, \araa, 55, 213, \dodoi{10.1146/annurev-astro-082214-122339}

\bibitem[{{Sobacchi} \& {Mesinger}(2013)}]{2013MNRAS.432.3340S}
{Sobacchi}, E., \& {Mesinger}, A. 2013, \mnras, 432, 3340,
  \dodoi{10.1093/mnras/stt693}

\bibitem[{Stefanon {et~al.}(2019)Stefanon, Labbe, Bouwens, Oesch, Ashby,
  Caputi, Franx, Fynbo, Illingworth, Le~Fevre, Marchesini, McCracken,
  Milvang-Jensen, Muzzin, \& van Dokkum}]{Stefanon2019}
Stefanon, M., Labbe, I., Bouwens, R.~J., {et~al.} 2019, \apj, 883, 99

\bibitem[{{Sun} \& {Furlanetto}(2016{\natexlab{a}})}]{Sun2016}
{Sun}, G., \& {Furlanetto}, S.~R. 2016{\natexlab{a}}, \mnras,
  \dodoi{10.1093/mnras/stw980}

\bibitem[{{Sun} \& {Furlanetto}(2016{\natexlab{b}})}]{SunFurlanetto16}
---. 2016{\natexlab{b}}, \mnras, 460, 417, \dodoi{10.1093/mnras/stw980}

\bibitem[{{Tacchella} {et~al.}(2018){Tacchella}, {Bose}, {Conroy},
  {Eisenstein}, \& {Johnson}}]{Tacchella2018}
{Tacchella}, S., {Bose}, S., {Conroy}, C., {Eisenstein}, D.~J., \& {Johnson},
  B.~D. 2018, \apj, 868, 92, \dodoi{10.3847/1538-4357/aae8e0}

\bibitem[{Tacchella {et~al.}(2013)Tacchella, Trenti, \&
  Carollo}]{Tacchella2013}
Tacchella, S., Trenti, M., \& Carollo, C.~M. 2013, \apjl, 768, L37

\bibitem[{Trenti {et~al.}(2010)Trenti, Stiavelli, Bouwens, Oesch, Shull,
  Illingworth, Bradley, \& Carollo}]{Trenti2010}
Trenti, M., Stiavelli, M., Bouwens, R.~J., {et~al.} 2010, The Astrophysical
  Journal, 714, L202

\bibitem[{{Villaescusa-Navarro} {et~al.}(2020){Villaescusa-Navarro},
  {Angl{\'e}s-Alc{\'a}zar}, {Genel}, {Spergel}, {Somerville}, {Dave},
  {Pillepich}, {Hernquist}, {Nelson}, {Torrey}, {Narayanan}, {Li}, {Philcox},
  {La Torre}, {Delgado}, {Ho}, {Hassan}, {Burkhart}, {Wadekar}, {Battaglia}, \&
  {Contardo}}]{2020arXiv201000619V}
{Villaescusa-Navarro}, F., {Angl{\'e}s-Alc{\'a}zar}, D., {Genel}, S., {et~al.}
  2020, arXiv e-prints, arXiv:2010.00619.
\newblock \doarXiv{2010.00619}

\bibitem[{{Virtanen} {et~al.}(2020){Virtanen}, {Gommers}, {Oliphant},
  {Haberland}, {Reddy}, {Cournapeau}, {Burovski}, {Peterson}, {Weckesser},
  {Bright}, {van der Walt}, {Brett}, {Wilson}, {Millman}, {Mayorov}, {Nelson},
  {Jones}, {Kern}, {Larson}, {Carey}, {Polat}, {Feng}, {Moore}, {VanderPlas},
  {Laxalde}, {Perktold}, {Cimrman}, {Henriksen}, {Quintero}, {Harris},
  {Archibald}, {Ribeiro}, {Pedregosa}, {van Mulbregt}, \& {SciPy 1. 0
  Contributors}}]{2020NatMe..17..261V}
{Virtanen}, P., {Gommers}, R., {Oliphant}, T.~E., {et~al.} 2020, Nature
  Methods, 17, 261, \dodoi{10.1038/s41592-019-0686-2}

\bibitem[{{Wang} {et~al.}(2020){Wang}, {Davies}, {Yang}, {Hennawi}, {Fan},
  {Barth}, {Jiang}, {Wu}, {Mudd}, {Ba{\~n}ados}, {Bian}, {Decarli}, {Eilers},
  {Farina}, {Venemans}, {Walter}, \& {Yue}}]{Wang20}
{Wang}, F., {Davies}, F.~B., {Yang}, J., {et~al.} 2020, \apj, 896, 23,
  \dodoi{10.3847/1538-4357/ab8c45}

\bibitem[{{Whitler} {et~al.}(2020){Whitler}, {Mason}, {Ren}, {Dijkstra},
  {Mesinger}, {Pentericci}, {Trenti}, \& {Treu}}]{2020MNRAS.495.3602W}
{Whitler}, L.~R., {Mason}, C.~A., {Ren}, K., {et~al.} 2020, \mnras, 495, 3602,
  \dodoi{10.1093/mnras/staa1178}

\bibitem[{{Yang} {et~al.}(2020){Yang}, {Wang}, {Fan}, {Hennawi}, {Davies},
  {Yue}, {Banados}, {Wu}, {Venemans}, {Barth}, {Bian}, {Boutsia}, {Decarli},
  {Farina}, {Green}, {Jiang}, {Li}, {Mazzucchelli}, \& {Walter}}]{Yang20}
{Yang}, J., {Wang}, F., {Fan}, X., {et~al.} 2020, \apjl, 897, L14,
  \dodoi{10.3847/2041-8213/ab9c26}

\bibitem[{{Yung} {et~al.}(2019){Yung}, {Somerville}, {Finkelstein}, {Popping},
  \& {Dav{\'e}}}]{Yung2019a}
{Yung}, L.~Y.~A., {Somerville}, R.~S., {Finkelstein}, S.~L., {Popping}, G., \&
  {Dav{\'e}}, R. 2019, \mnras, 483, 2983, \dodoi{10.1093/mnras/sty3241}

\bibitem[{{Yung} {et~al.}(2020){Yung}, {Somerville}, {Finkelstein}, {Popping},
  {Dav{\'e}}, {Venkatesan}, {Behroozi}, \& {Ferguson}}]{Yung20}
{Yung}, L.~Y.~A., {Somerville}, R.~S., {Finkelstein}, S.~L., {et~al.} 2020,
  \mnras, 496, 4574, \dodoi{10.1093/mnras/staa1800}

\bibitem[{{Zubeldia} \& {Challinor}(2019)}]{2019MNRAS.489..401Z}
{Zubeldia}, {\'I}., \& {Challinor}, A. 2019, \mnras, 489, 401,
  \dodoi{10.1093/mnras/stz2153}

\end{thebibliography}
 \newcommand{\noop}[1]{}



\end{document}